\documentclass[showpacs,aps,preprintnumbers,letterpaper]{revtex4-1}
\usepackage{amsmath,amssymb}
\usepackage{epsfig}
\usepackage{graphicx}
\usepackage{amsmath}
\usepackage{slashed}
\usepackage{amsfonts}
\usepackage{epstopdf}
\usepackage{color}
\usepackage[caption=false]{subfig}
\usepackage[pdftex, pdfborder= 0 0 0, citecolor=blue, urlcolor=blue, linkcolor=blue, colorlinks=true, bookmarksopen=true]{hyperref}

\newcommand{\be}{\begin{equation}}
\newcommand{\ee}{\end{equation}}
\newcommand{\bea}{\begin{eqnarray}}
\newcommand{\eea}{\end{eqnarray}}
\newcommand{\ba}{\begin{eqnarray}}
\newcommand{\ea}{\end{eqnarray}}

\begin{document}

\title{ Many-body forces and nucleon clustering near the QCD critical point }
\author{ Dallas DeMartini and
 Edward Shuryak }
 
\affiliation{ Department of Physics and Astronomy, \\ Stony Brook University,\\
Stony Brook, NY 11794, USA}

\begin{abstract} 
It has been proposed that one can look for the QCD critical point (CP) by the Beam Energy Scan (BES) accurately monitoring event-by-event
fluctuations. This experimental program is under way
at the BNL RHIC collider. Separately, it has been studied how
clustering of nucleons at freezeout affects proton multiplicity distribution and light nuclei production. 
It was found that even a minor increase of the range of nuclear forces dramatically increases clustering, while 
large correlation length $\xi$ near CP makes attraction due to binary forces unrealistically  large. In this paper we show that repulsive many-body forces near CP should overcome the
binary ones and effectively suppress clustering. We also
discuss current experimental data and point out locations
at which a certain drop in clustering may already be observed.
 \end{abstract}

\maketitle
\section{Introduction}
The original work  \cite{Stephanov:1998dy} proposed
a search for the (hypothetical) QCD critical point (CP), by measurements  
of the event-by-event fluctuations with the Beam Energy Scan (BES), currently
 adopted as one of  major programs of the BNL Relativistic Heavy Ion Collider. One part of it, called BES-I,
is by now completed, with BES-II --  involving even lower collision energies, combining collider and fixed
target modes -- are yet to be performed.

Before we go into details of the calculations, let us present qualitatively  the main idea of this paper. Suppose the CP indeed exists, and is located in the part of the phase diagram near the freezeout line of BES program. 
 Furthermore, while scanning
this line,  for some (yet unknown) beam energy
the freezeout conditions happens to reach 
maximal value of the correlation length  $\xi$,
as compared to all other collision energies.  
What are the  observables sensitive to $\xi$?
And, more specifically, to what scale of $\xi$ would they show observable signals? 

One possibility actively discussed  is related with
hydrodynamic  (sounds-like) fluctuations of the density,  with the wavelength
comparable to $\xi$. One is expecting their 
enhancement due   to critical opalescence near CP. 

In heavy-ion collisions -- due to unprecedented small viscosity -- we indeed can observe several harmonics
of sound. They are numerated by harmonic number $n$ in  azimuthal angle $\phi$.  The maximal harmonic number observed    is currently at $n_{max}=9$ (ALICE) at LHC
energies, while at BES energies  it is about 
$n_{max}=6$ (STAR). 
The dependence
of harmonic amplitude on $n$ is well explained by the so called 
``acoustic damping" \cite{Staig:2010pn,Lacey:2013is} 
according to which $A(n)\sim exp(- n^2 \eta * const)$
with $\eta$ being matter shear  viscosity. The maximal
$n$ corresponds to statistical noise and depends on
available number of events detected.

These harmonics correspond to sound propagation  along the fireball surface,  inducing  correlations in $\phi$ of
 secondaries emitted from this surface.
 The  maximal harmonic number $n_{max}=6$
corresponds to minimal sound wavelength
\begin{equation}
\lambda_{min}={2\pi R \over n_{max}} \sim 6\, fm 
\end{equation} 
where $R$ is the fireball radius.
Unfortunately, $\xi$ of such large scale  is 
 unlikely to be reached in the scan. Therefore, more sophisticated correlations would be needed, perhaps combining azimuthal and rapidity correlations, aiming at the yet unobserved tails of sounds.

 We propose another observable, sensitive to significantly smaller scale  
 $$\xi_{max}\sim 1.5-2 \, fm$$ 
According to \cite{Shuryak:2018lgd,Shuryak:2019ikv}, 
this is the natural scale of the size of  few-nucleon
correlations, called  $preclusters$.
Their existence is due to the ordinary nuclear forces,
and their experimental manifestations are: \\ 
(i) higher moments (e.g. kurtosis, the 4th moment) of the proton multiplicity distribution;\\
(ii) yields of light nuclei --  $d$, $t$, $^3He$, $^4He$
-- due to additional feed down from precluster decays.

\begin{figure}
	\centering
	\includegraphics[width=0.7\linewidth]{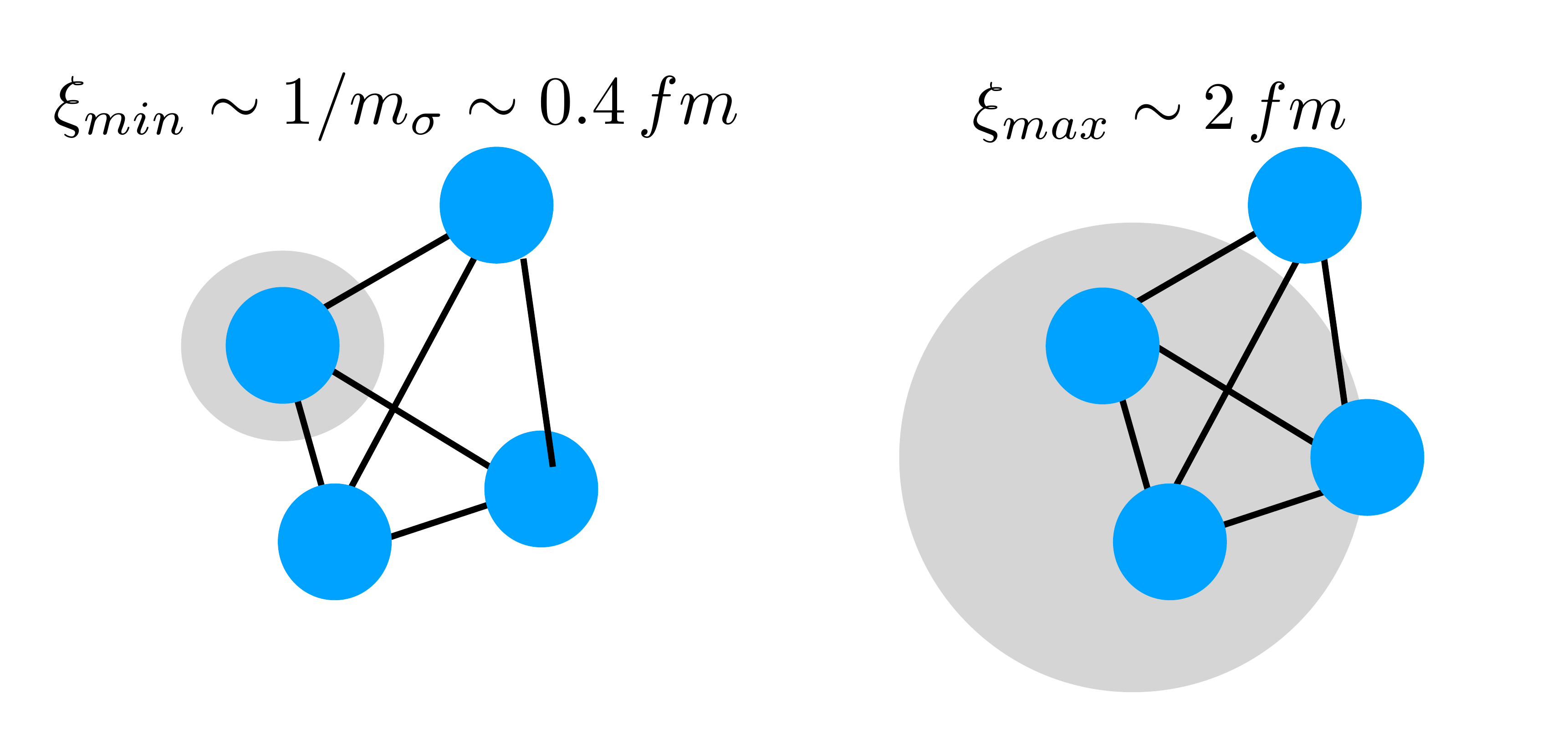}
	\caption{(Color online) Preclusters of four nucleons, shown by blue circles. Six lines connecting them indicate
		binary potentials.
		 The gray area indicates the range of forces between them, for standard nuclear forces (left) and near the critical point (right). 
	 In the latter case the  interaction is not only binary but many-body ones also appear. 
  }
	\label{fig:ximinmax}
\end{figure}

As we will show below, the interplay of attractive binary and repulsive  many-body forces is expected 
to show strong non-monotonous behavior of precluster
formation probability during the BES. 
The idea is illustrated in Fig. \ref{fig:ximinmax}.
The left one, 
 far from CP, has the usual short range ($\sim 1/m_\sigma\sim 0.4$ fm) of nuclear forces. Near CP, where
correlation length is of the order of cluster size  (Fig. \ref{fig:ximinmax} right) correlations of  nucleons in the cluster become stronger. 
As we will show, evaluating the magnitude and even the sign of the effect is rather nontrivial.

Now, with the main idea already spelled out,
let us introduce the subject more systematically.
While the shape and amplitude of critical fluctuations in the vicinity of CP are 
rather intricate, we do expect the  CP  to 
belong to the 3D Ising universality class, which has been
studied  for decades, analytically and numerically.
  One way to characterize fluctuations of the critical mode $\phi$ near the  CP is via the {\em cumulants of the critical field}
 \begin{equation} \kappa_2=\langle  \phi^2 \rangle, \,\,\,    \kappa_3=\langle  \phi^3 \rangle, \,\,\,    \kappa_4=\langle  \phi^4 \rangle -3 \langle  \phi^2 \rangle^2 \label{eqn_4moments}
 \end{equation} 
As   Stephanov \cite{1104.1627} pointed out,  such
cumulants can be related to certain diagrams,
containing higher powers of the correlation length $\xi$
and coupling constants, from the effective action describing the fluctuations. 

 Unfortunately, we do not have any experimental means to directly 
 access fluctuations of the critical mode $\phi$. Since 
 it is  expected that it couples to pions rather weakly,  naturally it was suggested to use the only
 other species copiously produced, namely nucleons.

In Refs  \cite{Stephanov:1998dy,1104.1627} moments 
of the critical field fluctuations (\ref{eqn_4moments}) were related to those
of
 the nucleons,  under  crucial assumption that
 {\em nucleons are uncorrelated }  by any other effects.
 If locations of the
nucleons  can  be integrated independently, each external line of these diagrams becomes simply
a propagator integrated over  space, namely
$$ \int d^3 r {exp(-r/\xi) \over 4\pi r} =\xi^2.$$
    Unfortunately, this simplifying assumption 
is incorrect in reality. Conventional nuclear forces do
create significant correlations between them. 
Rather nontrivially, they survive 
 even at the freezeout stage of heavy ion collisions,
with temperature $T\sim 100\, MeV$ much larger than
conventional bindings of light nuclei.    As shown  
in Refs. \cite{Shuryak:2018lgd,Shuryak:2019ikv},  there  exist phenomenon of nucleon $preclustering$, starting
from four-nucleon systems. One needs six (or more) pair potentials for correlations to remain appreciable at these high temperatures.  

 Preclustering phenomena were
 studied  by a number of theoretical 
tools: \\
(i) classical molecular dynamics \cite{Shuryak:2018lgd},\\
(ii) semiclassical ``flucton" 
method at finite temperatures \cite{Shuryak:2019ikv}; \\
(iii) quantum mechanics in hyperspherical coordinates \cite{Shuryak:2019ikv};\\
(iv) the (first principle) path-integral Monte Carlo (PIMC) \cite{DeMartini:2020hka}.

\begin{figure}[b]
\begin{center}
\includegraphics[width=8cm]{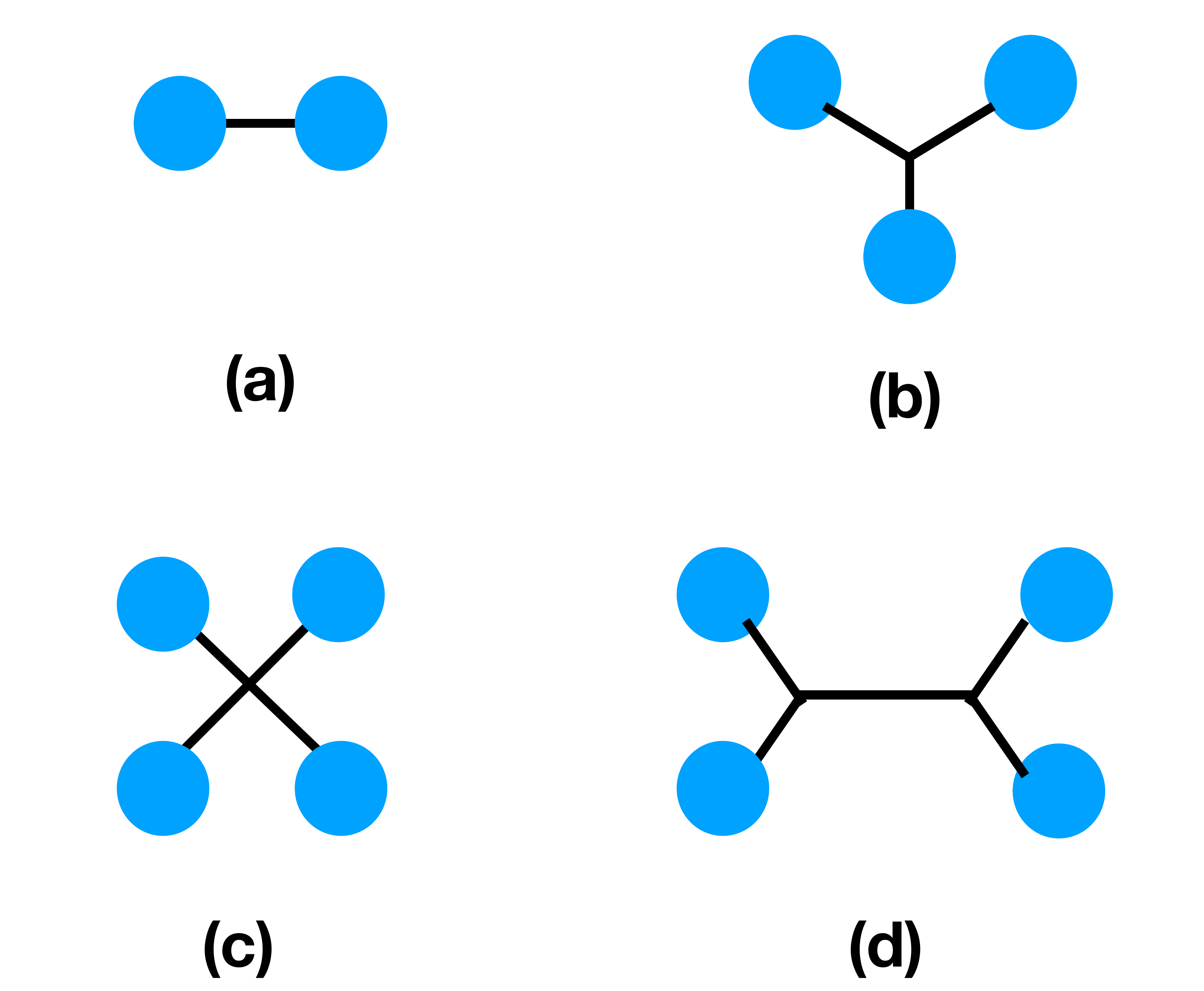}
\caption{(Color online) Diagrams representing the many-body interactions of the four-nucleon cluster. Blue circles are nucleons, black lines are propagators of the $\phi$ fields}
\label{fig_diag}
\end{center}
\end{figure}

We will use some results of our previous paper on the subject \cite{DeMartini:2020hka} based on PIMC simulations at appropriate temperatures and densities
of BES freezeouts. For four-nucleon clusters we
calculated the 9-dimensional effective volume of the precluster, entering the 4th-order virial coefficient. We have shown that while precluster phenomenon only contribute to 
multiplicity at a sub-percent level, its positive contribution to $kurtosis$ of the proton multiplicity distribution
becomes of order one for collision energies at and below $\sqrt{s}=7.7$ GeV, as it is indeed
observed by STAR collaboration. 

While in  all these papers \cite{Shuryak:2018lgd,Shuryak:2019ikv,DeMartini:2020hka}
a variety of theoretical tools were used, the emphasis
was on their consistency. Therefore the same 
  $binary$ nuclear forces -- the simplified Walecka model -- were used in all of them. The issues of CP
 were addressed only peripherally, by binary forces 
modified by added exchanges of longer-range critical mode. Since the effect of that was persistently found to be catastrophic, 
 it was clear   that this approach could not  possibly be an accurate description of the interactions near CP.

And indeed, as we will show 
in this paper, only with the inclusion of $many-body$ forces induced by critical fluctuations near  the  hypothetical CP
resolves the puzzle. Furthermore, with
 presumed growth of the  correlation length $\xi$, repulsive three and four-nucleon forces grow 
 $stronger$ than binary ones, reversing the dependence on $\xi$. Basically, we will show that
 all preclustering
  should be suppressed in a small vicinity of CP. Thus, our calculations indeed predict strong $non-monotonous$ signal for BES,
 starting as an enhancement of clustering, to its full absence near the CP, and then back to enhancement at the other side of the CP.

(Before we begin our discussion, let us state for clarity 
that in this paper we are $not$ interested in the most generic problem
of the many-body forces influencing the thermodynamics of
infinite matter (at freezeout). Traditional studies of nuclear matter do include well documented  three-body forces,
derived from precise treatment of light nuclei. 
Those are not important here, since
the  nucleon density at freezeout conditions  of heavy ion collisions of interest are even smaller than 
nuclear matter density.  Also, as one can see below,
the effects we discuss are much larger than those 3-body forces.)

Note that we discuss four-nucleon clusters with specific
 flavor-spin arrangement $p^\uparrow p^ \downarrow n^\uparrow n^\downarrow$, with all four nucleons 
 being distinguishable particles, so Pauli blocking is completely absent.
 This also simplifies combinatorial factors and reduced the technical challenges of the previous PIMC calculations. 

The structure of the paper is as follows: in section \ref{sec_diags} we introduce some lowest-order diagrams describing the
 interaction of the critical mode with nucleons and with itself, and qualitatively discuss their signs and magnitudes. Dependence of the diagram magnitude on the
 cluster  size relative to the correlation length is discussed in section
 \ref{sec_sizes}. In the next section \ref{sec_shapes} we average the diagrams  over cluster shapes, using snapshots from
 the PIMC performed in Ref. \cite{DeMartini:2020hka}.
In section \ref{sec_potential} we discuss the universal
effective potential $\Omega(\phi)$ describing
critical fluctuations on the critical line of Ising-class
phase transitions. In section  \ref{sec_deformed} we consider a deformation of this potential by some external current $J$, shifting
a bit from the critical line, and representing the
freezeout path on the QCD phase diagram. Nonlinear
coefficients of these deformed potentials  are used
as coupling constants in many-body diagrams.  Combining
those with the calculations of the diagrams themselves, we get to the results shown in Fig. \ref{fig:expabc}. According to it, strong attraction due to exchange of the critical mode between the nucleons enhances clustering, with maximum at $t\approx 0.2$, and at smaller $t<0.11$ (closer to CP) it changes to repulsion,
soon suppressing clustering. In section \ref{sec_discussion} we summarize the paper and discuss current status of relevant experimental observables.

\section{Three and four-nucleon forces and the  four-nucleon clusters} \label{sec_diags}
\subsection{Effect of critical binary potential}
Refs. \cite{Shuryak:2018lgd,Shuryak:2019ikv,DeMartini:2020hka} all discussed
the effect of the the hypothetical critical point on nucleon interactions, but only via $binary$ forces.
 The critical fluctuations were assumed to add to conventional nuclear force a new binary potential corresponding to the diagram Fig. \ref{fig_diag}(a)
\be  V_a=-g_c^2 \langle  \phi(\vec r) \phi(0) \rangle=-{g_c^2\over 4\pi} {exp(-r/\xi) \over r} \ee
Since this potential was included in the exponential of the action, all of its iterations
were also included. 
The coupling of the critical mode to nucleons $g_c$ of course depends on 
the nature of the critical mode $\phi$. While in principle it can be estimated from
mapping of Ising coordinates to QCD phase diagram, it does not belong to a
class of observables uniquely predicted by universality arguments.  One perhaps can view
$\phi$ as having some admixture of the lowest (isoscalar) mesons $\sigma,\omega$,
(or more precisely,  the lowest-mass edge of the corresponding spectral densities). But, since
the couplings to them have opposite sign, the magnitude of $g_c$ is hard to estimate, and we will use it as a free parameter.  
 
As shown in all these works \cite{Shuryak:2018lgd,Shuryak:2019ikv,DeMartini:2020hka},  such approach leads to huge effects, which were judged to be unrealistic. Indeed, if the correlation length grows to $\xi>  2 \,fm\sim 1/(100\, MeV)$, all six pair terms in a four-nucleon cluster are comparable, leading to large correlation $\sim exp(6 |V_a|/T)$.   

In fact, it has been noticed previously by one of us \cite{Shuryak:2006eb} that such approach would lead to
catastrophic phenomena when $\xi \rightarrow \infty$. Indeed, in this limit we will have attractive
Newton-like potential between all nucleons in the fireball acting $coherently$. Since the total number
of nucleons in the fireball is $N=O(100)$, the number of  pairs $N(N-1)/2$ is so huge that  for any meaningful $g_c$ 
(larger than QED electric coupling)
one faces a (gravitation-style) collapse of the system! 
Looking for effects which can prevent this from happening, one naturally should consider the multi-nucleon forces.  

\subsection{Qualitative discussion of the multibody effects}
Before we discuss  phenomena associated with the critical point, let us recall how
the usual nuclear potential and related clustering enter
the thermodynamics. As explained in detail in our previous work 
\cite{DeMartini:2020hka}, the 4-body clusters made of 4 distinguishable nucleons contribute
the potential energy part of the statistical sum in the form of the fourth virial coefficient.

The potential part of the partition function (of a single species system) of $N$ particles  can be re-written in the form 
\begin{equation} Z_{pot}=1+{1 \over V^N} \int d^3 x_1...\int d^3 x_N \big[ e^{\big(-\sum_{i>j} V(\vec x_i -\vec x_j)/T\big) } -1\ \big] 
\end{equation}
by adding and subtracting 1. Since we focus on clusters of distinguishable 4 particles, coordinates of all others can be integrated out, as well as the coordinates of
its center of mass. What is left is
\begin{equation}  Z_{pot}=1+\big({N\over  4 }\big)^4({V_{cor} \over V^3}) 
\end{equation}
where the so-called 9-dimensional  correlation volume is
\begin{equation}
V_{cor}^{(9)} = \frac{32}{105}\pi^4 \int d\rho \rho^8 (P(\rho)-1).
\end{equation}
Here $P(\rho)$ is the probability distribution in the 9-dimensional hyperdistance $\rho$ normalized to that of a non-interacting ideal gas, 
and the factor in front is the solid angle in 9 dimensions. We neglect repulsion and integrate over the region in which the integrand is positive. 
The addition to the free energy is then
$ \Delta (-T log(Z))= - T n^3 V^{(9)}_{cor}{N \over 4^4} $, same as to the grand partition sum. Differentiating it with respect to $\mu$, present in each $N$,
one finds the addition to particle number $ \Delta N/N=  n^3 V^{(9)}_{cor}/4^3 $ .

The magnitude of this effective volume depends on 
the temperature and density of the matter, and it was calculated in our PIMC simulations \cite{DeMartini:2020hka}. For example, at kinetic freezeout conditions of $\sqrt{s}=7.7$ GeV, we found 
\be
V_{cor}^{(9)}(7.7)\approx 4.3\cdot 10^4\, fm^9
\ee
To put it in proper prospective, one can define the ``density of the cluster" as 
\be n_{cl}\equiv {4\over\big( V_{cor}^{(9)}\big)^{1/3} }
\ee
which for  $\sqrt{s}=7.7$ GeV is  $n_{cl} \approx 0.114\, /fm^3$. This value is about 3 times the density of 
ambient matter $n_B(7.7)\approx 0.037 \, /fm^3$.

In our previous work, the PIMC action included only the binary forces between nucleons, either the standard ones (simplified to the Walecka form),
or modified due to chiral crossover via reduced sigma mass. In this work our task is to include the 
many-body forces appearing near the hypothetical critical point.

Since below we will need to compare the inter-nucleon separations to
the critical correlation length $\xi$, we will also 
define it by a cubic root of the respective densities 
\be R_{amb}\equiv n_B^{-1/3}\approx 3.0\, fm, \,\,\,\,\,\, R_{cl}\equiv n_{cl}^{-1/3}\approx 2.0\, fm 
\ee

 The difference between these values may not appear to be large, but it would turn out to be crucial, as it will enter
the relevant formulae in large powers. We do not yet know if the CP
exists or not on the phase diagram, and we do not know what magnitude its maximal correlation length $\xi$ may reach on the freezeout line. For estimates we will  assume that
$ \xi_{max} \sim 2\, fm $ can be reached, the value
 comparable to $R_{cl}$ defined above.  As we will see, at such value the multibody forces
 are important for clusters but $not$ for ambient matter.

Let us now approach the critical point effects,  using first the simplest approach available,
known as Landau's mean-field model. We also assume, for simplicity, that the freezeout and crossover transition line coincide.  If so, the effective potential has $\phi \rightarrow -\phi$ symmetry and therefore odd powers of it
must vanish, $\lambda_3=0$ and with it $V_b,V_d=0$ ($V_i$ are the interactions of diagram (i)). Traditionally
the Landau potential has only the mass term and 
nonzero 4-point vertex coupling $\lambda_4$.
(Yes, we know the Landau potential does not correspond to CP, and nowadays is only used as the initial
conditions for RG flow calculations. We will discuss
proper critical potential below.)

This approximation leaves us with only two terms: the attractive two-body term $V_a\sim n_B^2$
and the repulsive four-body term $V_c\sim \lambda_4 n_B^4 $. At the small density
of ambient matter, $n_B$ is small and the former dominates,
while at the high density of the cluster, the latter dominates. 

The free energy per particle  is 
\begin{equation} {F \over N}\sim  -{g_c^2 \over R}\big({\xi \over R}\big)^2+ {\lambda_4 g_c^4\over R}  \big({\xi \over R}\big)^8. \end{equation}
In an Ising-type critical point in fact the quartic coupling vanishes, as $\lambda_4\sim 1/\xi$, making the effective power
of it in the last term seven, not eight. Still, the 
dependence on $\xi$ is the same: negative at small $\xi$ 
is reversed to large and positive as $\xi$ grows. This means
CP should {\em suppress preclustering} and thus reduce feed-down from the $4N$ system!

The magnitude of the couplings $g_c$, $\lambda_4$ are not yet
known, but the effects of the $\xi/R$ ratios can be calculated.
While in clusters this ratio is just about 1, with all its powers, 
for ambient matter these two terms have them be equal to
\be \big({\xi_{max}\over R_{amb}}\big)^2\approx 0.444, \,\,\, 
 \big({\xi_{max}\over R_{amb}}\big)^7\approx 0.058
 \ee
 and the many-body repulsion term is relatively small.

The critical fluctuation effects
thus can work $against$ clustering, reducing the cluster volume $V_{cor}^{(9)}$, and thus leading to a $reduction$ of the kurtosis. Note,
that this approximation  corresponds to approaching the CP from
smaller to large density, or $\mu_B$, or approaching with collisions at  energies $above$ that of CP. 

\subsection{Multibody forces in four-nucleon clusters}

In general, the potential part of the partition function should include both binary and many-body
forces. While the former ones were included 
in PIMC simulations, the latter were not there.
Our task in this work to do so, in particular
for the many-body forces appearing due to nonlinear
effective Lagrangian of the critical mode. 

Let us introduce the notations we use. For three-body forces induced by diagram (b) we define function

\be V_b\big(\vec x_1,\vec x_2,\vec x_3\big)\equiv \int d^3 u   D(\vec x_1-\vec u) D(\vec x_2-\vec u) D(\vec x_3-\vec u)\ee
 
where
 \be D(r)=exp(-r/\xi)/r \ee is the binary Yukawa potential.  Note that this function is dimensionless, and that
 we do not include here the factor $1/4\pi$ present in 3d  propagator, which will be included later with the couplings.

Similarly, we define  four-body function for diagram (c), we have

\be V_c\big(\vec x_1,\vec x_2,\vec x_3,\vec x_4\big)\equiv \int d^3 u   D(\vec x_1-\vec u) D(\vec x_2-\vec u) D(\vec x_3-\vec u)D(\vec x_3-\vec u).\ee
Note that its dimension will be $[fm^{-1}]$.

Finally, for diagram (d) we define

\be V_d\big(\vec x_1,\vec x_2,\vec x_3,\vec x_4\big)\equiv \int d^3 u d^3v D(\vec x_1-\vec u)D(\vec x_2-\vec u)D(\vec u-\vec v)D(\vec x_3-\vec v)D(\vec x_4-\vec v)
\ee 
with corresponding dimension $[fm]$.

 These functions depend on the coordinates of 3 or 4 nucleons, and should
 be averaged over many-body density matrix $P\big(\vec x_1,\vec x_2,\vec x_3,\vec x_4\big)$ of the clusters.

 Using these definitions, we can write the effective potential for four-nucleon cluster  in the following form
 
\begin{eqnarray}
V_{abcd}=&&-{4\cdot 3 \over 2} {g_c^2\over 4\pi}{exp{(- r_{ij}/\xi)}\over  r_{ij}}  +4 \cdot 3!\lambda_3({g_c \over 4\pi})^3 V_b\nonumber  \\
&& +4!\lambda_4 ({g_c \over 4\pi})^4 V_c - 4!{\lambda_3^2\over 8\pi}({g_c \over 4\pi})^4 V_d,
\end{eqnarray}
where we have now restored combinatorial factors and signs. All interactions $V_i$ generically depend on all nine hypercoordinates, although we will make simplifying assumptions later. Note that
an extra $1/8\pi$ in the last term comes from 
$1/2!$ of the second order expansion and $1/4\pi$
from an extra
intermediate propagator between the vertices.

Generally speaking, this many-body potential should
be included in PIMC simulations, as it was done with the binary potential, to directly observe its effect on clustering. It is not however practical 
to do so as they include extra multidimensional integrations over the
locations of the nonlinear vertices, and are thus too computationally intensive at present.   

We therefore adopt the {\em perturbative approach}, in which 
all locations of the nucleons $\vec x_i$ 
in the cluster  are to be  averaged
over the appropriate 9-dimensional density matrix
calculated in PIMC with {\em the binary interactions
only }.

 In doing this average, we would like to separate the dependencies on the ``hyperdistance" $\rho$ and the ``shapes" (angular variables) of the cluster. The former is defined via most-symmetric definition of the hyperdistance $\rho$ (\ref{rho_definition}) coordinate.  

\subsection{Dependence of multibody forces on the cluster shape and the correlation length} \label{sec_sizes}

As a warm-up, we calculate the diagrams for two specific shapes. The most symmetric one is a $tetrahedral$ shape, in which
all pair distances are the same $L_{tet}$. Another shape we considered is a {\em flat square} with size $L_{sq}$: in order for both to correspond to the same
 hyperdistance $\rho$, they should be related by
 \be 6 L_{tet}^2=(4+2\cdot 2 )L_{sq}^2 = 4\rho^2 \ee

 \begin{figure}[h]
	 \includegraphics[width=1.0\linewidth]{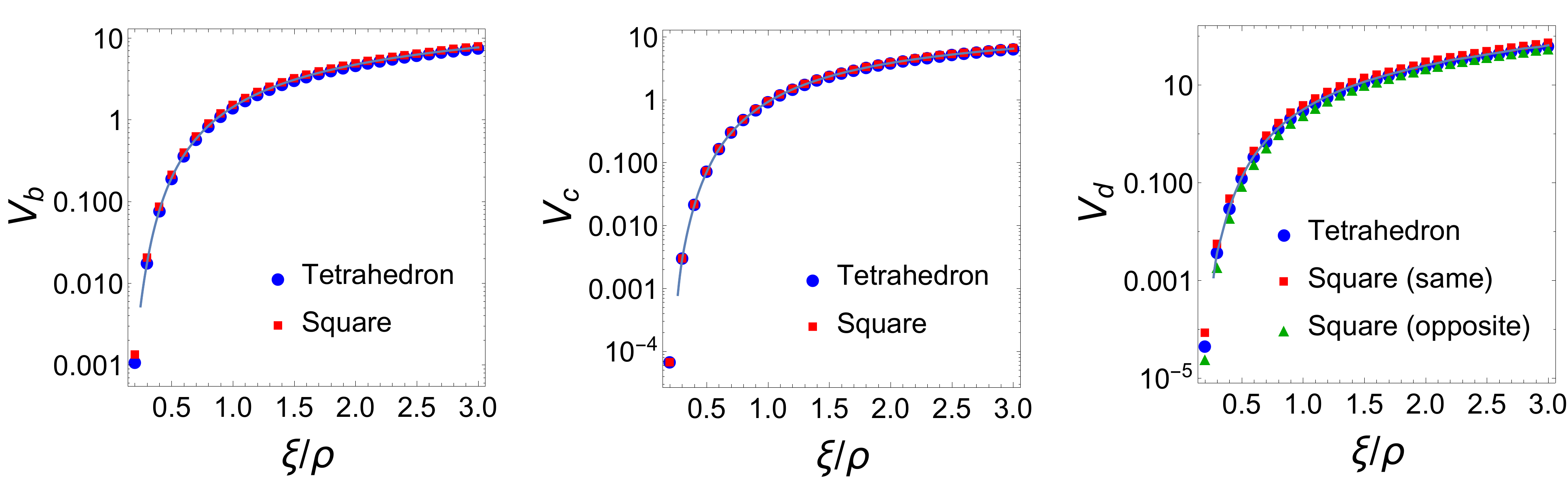}
 \caption{
 (Color online) Interactions $V_b$ (left), $V_c$ [fm$^{-1}$] (center), and $V_d$ [fm] (right) corresponding to diagrams (b,c,d) of Fig. \ref{fig_diag}, respectively, as a function of the correlation-length-to-hyperdistance ratio $\xi/\rho$ for both the tetrahedral and square configurations. The curve is an interpolation of the tetrahedral data points. The distinction between the 'same' and 'opposite' square configurations for diagram (d) is explained in the text.    	
} 	\label{fig:v3}
 \end{figure}

The results are shown in Fig. \ref{fig:v3} as a function of the basic ratio $\xi/\rho$, and also in Table 1 for  $\xi/\rho=1$
. In diagram (d) there are two vertices and the square configuration can be further divided into two more configurations: one in which nucleons on the \textit{same} side of the square are connected to the same vertex and one in which nucleons on \textit{opposite} corners are connected. The same distinction can be made for the binary interaction (diagram (a)), where two nucleons on the \textit{same} or \textit{opposite} side of the square can be connected by the propagator.   
While $V_b$ and $V_c$ show very small dependence on cluster shape, it is not so for $V_d$. Both the tetrahedron and square are very symmetric configurations and we know from the previous work that even in the correlated cluster, there are not significant angular correlations between the nucleons. \\

Using these results and assuming, for simplicity,
a Landau form of effective action, with
only diagrams (a) and (c) included, one can access
the dependence of the cluster potential on the magnitude of the correlation length $\xi$.
Assuming further that all clusters have the same tetrahedral shapes, we define the average potential as
\begin{eqnarray} \label{eqn_energy_LG}
V_{tet}=-6 {g_c^2\over 4\pi} \langle V_a  \rangle   +4!\lambda_4 ({g_c \over 4\pi})^4 \langle V_c  \rangle  
\end{eqnarray}
Now we need to select reasonable values for the couplings. Some guidance on  the magnitude of $g$,
the coupling of $\phi$ to the nucleon, can be obtained from the Walecka model applications to nuclear matter. In it the sigma and omega couplings are 
\begin{equation} \label{eqn_couplings_walecka}
{g_\sigma^2 \over 4 \pi} = 6.04, \,\,\,\,
{g_\omega^2 \over 4 \pi} = 15.17.
\end{equation}
The critical mode $\phi$ is presumably some superposition of 
the (lowest-momenta parts of the spectral densities) with $\sigma,\omega$ quantum numbers. So, its coupling must be comparable. As a guess, in literature some round
intermediate number 
\begin{equation}
{g_c^2 \over 4 \pi}=10 
\end{equation}
was used, and
we take this value in estimates to follow.

 \begin{figure}[t]
 	\centering
\includegraphics[width=0.45\linewidth]{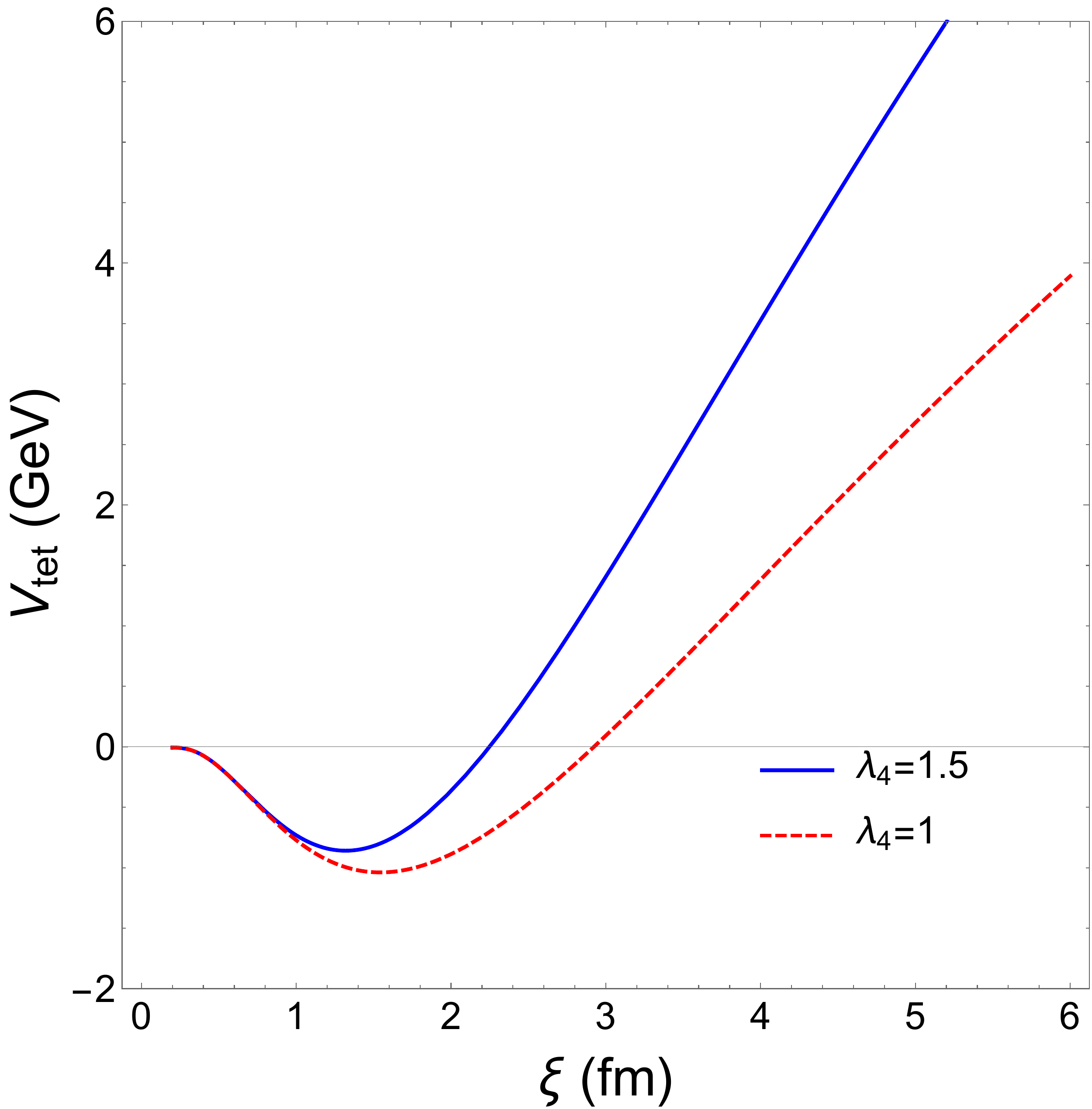}
 	\caption{
 		(Color online) Energy of four-nucleon tetrahedral
 		cluster $V_{tet}$ of size $\rho=2$ fm as a function of 
 		correlation length $\xi$. The 
 		critical mode-nucleon coupling is taken to be equal to nucleon-sigma meson coupling of the Walecka model	(\ref{eqn_couplings_walecka}), and two values of the four-point coupling $\lambda_4$ are used.
 	}
 	\label{fig:tet_of_xi}	
 \end{figure}

The value of quartic coupling 
$\lambda_4$ in the  Landau  model remains an  arbitrary parameter. So in Fig. \ref{fig:tet_of_xi} we show 
the dependence of the additional cluster energy (\ref{eqn_energy_LG}) as a function of the correlation length,
for its  two  values.
Naturally, at small $\xi$ all forces are very short range and additional energy is very small. With  $\xi$
growing to about 1.5 fm the six attractive potentials reach together a value of the order of $-1$ GeV,
but for larger $\xi$ values the quartic term mitigates attraction and turns the curve upward, eventually making
this additional energy positive. A similar trend would be seen for any other cluster shape. This
provides some initial understanding of possible role of the many-body forces.

\subsection{Averaging the multi-body forces over PIMC clusters}\label{sec_shapes}

\begin{figure}
 	\includegraphics[width=1.0\linewidth]{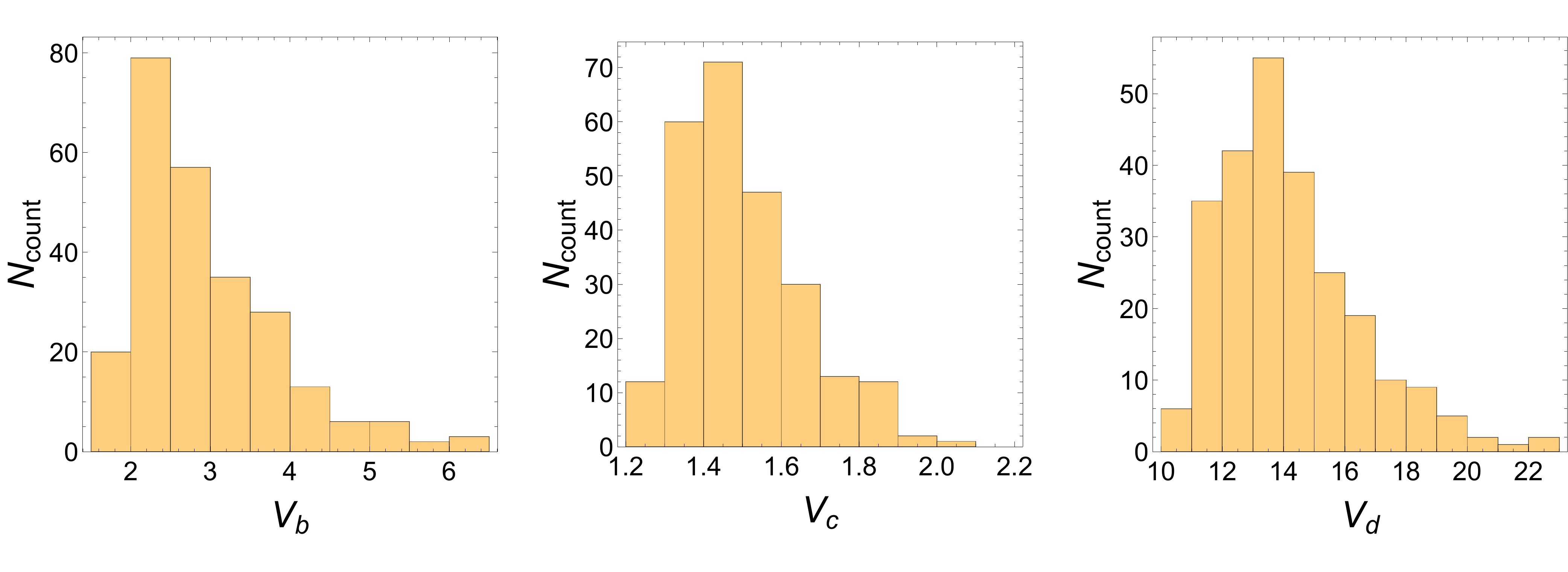}
 	\caption{
	 	Distribution of values of the multibody interactions $V_b$ (left), $V_c$ (center), and $V_d$ (right) corresponding to diagrams (b,c,d) of Fig. \ref{fig_diag}, respectively, in 250 configurations generated in PIMC simulation. All configuration have $1.49 < \rho < 1.51$ (fm) and were computed with $\xi = 2$ fm.   	
 	} 	\label{fig:hist1}
 \end{figure}

To make the analysis of the previous section a bit more quantitative, one needs to understand
the effect of averaging over all cluster shapes.
In order to  do so  we use the 9-dimensional configurations taken from PIMC simulation at a fixed value of $\rho$
and calculates all 4 diagrams for them. The distribution of values 
is shown in Fig. \ref{fig:hist1}. One finds that in fact there
where a wider distribution of values is seen than for the fixed shapes discussed before. For all three diagrams, variation by $\sim 50\%$ is seen from changing the shape but keeping $\rho$ fixed. This indicates that an accurate parameterization of these interactions requires not just dependence on hyperdistance $\rho$, but rather they must depend on the full set of $9$-dimensional hypercoordinates. However, we find that there is overall less sensitivity to system shape than to other quantities in the energy of the cluster, such as the nucleon-critical mode coupling $g_c$ or the external current $J$ (see Section \ref{sec_deformed}). Because the dependence on cluster shape is rather weak, we assume that the average interactions over all shapes is equal to that of the tetrahedral cluster of the same size, $\langle V_i(\xi/\rho) \rangle = V_{i,tet}(\xi/\rho)$.    

 After these preliminary studies of the diagrams for clusters of particular shapes,
 we perform the actual density matrix from our PIMC ensemble \cite{DeMartini:2020hka}.
 The results are shown as histograms of the values of the potentials $V_i$,
 for configurations. The \textit{cluster} configurations are chosen from those with $\rho < 3$ fm, the approximate maximum size of the cluster. \textit{Ambient} configurations are then chosen from those with $\rho > 3$ fm, where the average inter-nucleon binary interaction is small $\langle V_{NN} \rangle \simeq 0$ and no correlation is observed. For each set 5000 configurations are chosen from the PIMC simulation corresponding to conditions of kinetic freezeout at $\sqrt{s} = 7.7$ GeV. As expected, these values are quite different for these two subsets, indicating that 
 many-body forces are much more important within the clusters than for random (uncorrelated)
 nucleons.
 
  \begin{figure}
  	\includegraphics[width=1.0\linewidth]{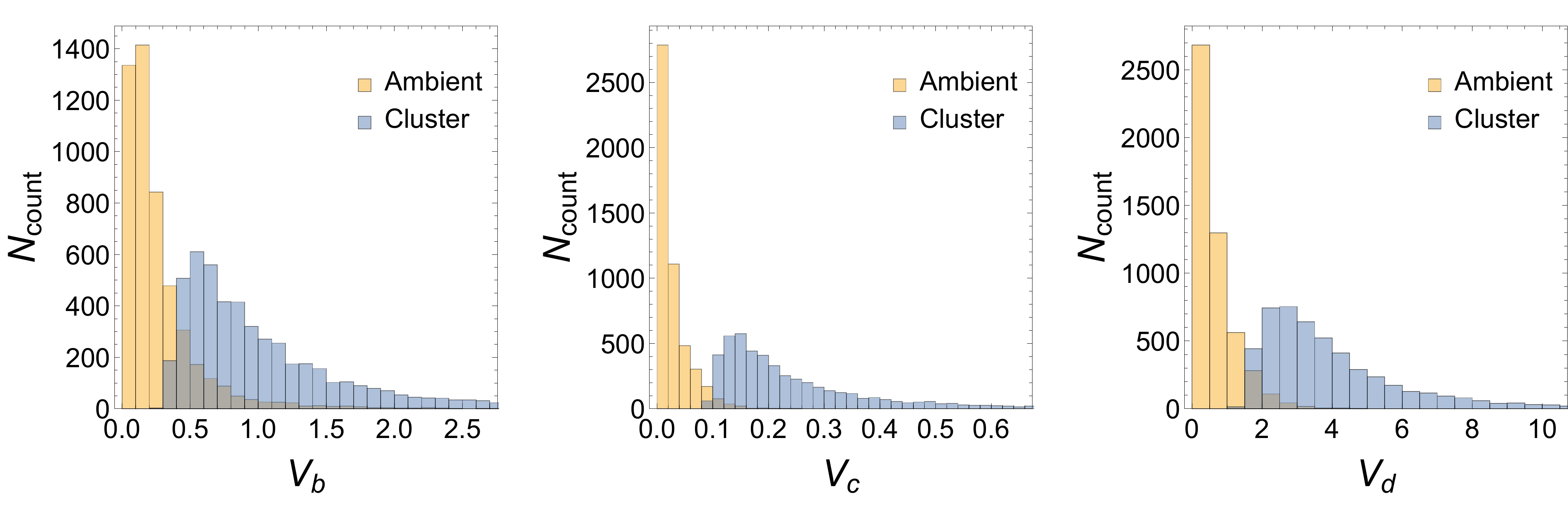}
  	\caption{
  		(Color online) Distribution of values of the multibody interactions $V_b$ (left), $V_c$ (center), and $V_d$ (right) corresponding to diagrams (b,c,d) of Fig. \ref{fig_diag}, respectively, in 5000 configurations each for the cluster ($\rho<3$ fm) and ambient nucleon matter ($\rho>3$ fm) generated in PIMC simulation. Calculation performed with $\xi = 2$ fm.   	
  	} 	\label{fig:amb}
  \end{figure}
 
 The results are plotted at Fig. \ref{fig:amb}. These histograms show that the many-body interactions are much stronger in the cluster compared to the ambient matter at freezeout. The distributions in the cluster possess both larger average values of the interactions and much longer high-value tails than the ambient matter distributions. The long tails of these distributions correspond to the 
 most compact clusters (with the smallest values of $\rho$).

 Comparing the average values of the interactions, one finds
  \be \frac{\langle V_a \rangle _{cl}}{\langle V_a \rangle _{amb}} = 2.63, \,\,\, \frac{\langle V_b \rangle _{cl}}{\langle V_b \rangle _{amb}} = 4.00, \,\,\, \frac{\langle V_c \rangle _{cl}}{\langle V_c \rangle _{amb}} = 10.73, \,\,\, \frac{\langle V_d \rangle _{cl}}{\langle V_d \rangle _{amb}} = 6.52. \ee 
    As expected, there is a clear hierarchy in these ratios. The dependence of the $N$-body diagrams on the ratio $\xi/\rho$ grows with $N$. Thus, the many-body interactions grow in their importance relative to the standard binary interaction as $\xi$ increases near CP. These ratios should grow at smaller values of row such as $\rho \sim 1.5$ fm, where peak spatial correlation is observed.

 \begin{table}
 	\caption{The first two rows are the average values of the diagrams in both the cluster and ambient nucleon matter computed with $\xi = 2$ fm. Latter three rows are the values of the diagrams for the specific geometries calculated with $\xi/\rho = 1$. All values are given without
 		couplings, combinatorial factors, signs, and factors of $4\pi$ in propagators: e.g.
 		 $V_a$ are given  without $-g_c^2/4\pi$.}
\begin{tabular}{|c|c|c|c| c |}
	\hline
	& a	& b & c & d \\
	\hline
$\langle V \rangle _{cl}$	& 0.110 & 1.092  & 0.292  & 4.184 \\
	\hline
$\langle V \rangle _{amb}$	& 0.289 & 0.273 & 0.027 & 0.642 \\
	\hline
$ V _{tet}$	& 0.541 & 1.434 & 0.924 & 3.143 \\
\hline
$ V  _{sq,same}$	& 0.697 & 1.485 & 0.956 &  3.713 \\
\hline
$ V _{sq,opp}$	& 0.368 & - &  - & 2.523 \\
\hline	
\end{tabular}
\end{table}

\section{The universal effective action for Ising-type critical fluctuations } \label{sec_potential}

The Landau model, used as an initial approximation, does $not$
however represent correct behavior near Ising-like critical points. Wilson's  expansion  in $\epsilon=4-d$ 
($d$ is space dimension) has found that under the
renormalization group flow
the Landau model goes into the fixed-point regime in 
 infrared, with small coupling  $\sim \epsilon$.  
While Wilson
famously calculated approximate values of the critical indices for $d=3$, further series in $\epsilon$ 
do not show good convergence and led to doubts about its accuracy at $\epsilon=1,d=3$.  

Exact renormalization group equations
were derived, using Wetterich exact RG equations,
and its solution for $d=3$ were worked out, for recent
reviews see Refs.
\cite{Berges:2000ew,Dupuis:2020fhh}. Unfortunately, obtaining the near-fixed-point solution can not be done analytically,
and therefore one relies on certain fits.

We will also use certain simplifying  approximations.  We will ignore renormalization of the propagator and its
index $\eta$, putting it to zero. So,  the kinetic 
term will be kept in its initial form $(\partial\phi)^2/2$, and
 the propagators  will be kept in their Yukawa form.

The effective vertices (powers of $\phi$ larger than 2) 
we get from  local form of 
the effective action $\Omega(\phi)$, will be obtained  
from fluctuation potential for
homogeneous constant fields $\phi(x)\rightarrow \langle \phi \rangle$.
The partition
  function  in the $x$-independent form  is just
\begin{equation}
 Z(J)=\int D\phi e^{(-\Omega(\phi)+J\phi)V_3/T},
\end{equation}  
where  $V_3$ is the volume of the system and $T$ is temperature, and the functional form of $\Omega(\phi)$
can be deduced from dependence on the external current $J$. 

We start with brief qualitative discussion
of possible form of this effective potential $\Omega(\phi)$ near CP, mentioning few  ``folklore"  arguments
suggesting that   $\Omega$ is effectively given by
a polynomial  of order 6. If one ``probes" $\Omega(\phi)$ by a nonzero external term $J$, the
 mean ``magnetization" $\langle \phi \rangle(J)$ index $\delta$ is defined as
\begin{equation} \langle \phi \rangle(J)\sim J^{1/\delta},\,\,\, \delta={d+2-\eta \over d-2+\eta }\approx 4.78.
\end{equation}
The number on the r.h.s. is empirical, from real and numerical experiments for various systems belonging to Ising universality class.

The minimum of the  potential shifted by $J\neq0 $
 is given by the solution of
\begin{equation}{d\Omega \over d\phi}=J.\end{equation} 
For $m\rightarrow 0$ and Landau theory, when
the only nonlinear term is $\phi^4$, one finds $\delta=3$, which is not close to the true value. 
The closest integer to 4.78 is 5. (Note that it corresponds to neglecting the $\eta\rightarrow 0$ 
 in general expression above, which we assumed anyway
 for propagators.) If so, 
it implies that $\Omega\sim \phi^6$.  One can therefore think 
 that a potential being a polynomial   of order six
 would be a good approximation to reality.
 
 The second argument is theoretical:  
including a $\phi^6$ term -- but not higher powers -- can be justified because this term is the last
 renormalizable  one, in $d=3$ space.

 There are of course multiple numerical studies of the Ising model 
 suggesting various fits of  $\Omega(\phi)$, at many lattices and $J$ values. 
  In particular,  good quality fits  
were reached in Ref. \cite{hep-lat/9401034},
 after the pre-exponent factor $\sqrt{d^2V/d\phi^2}$ was included. We will  follow 
  this paper, in which
\be \Omega(\phi) =\int d^3x \big[{(\partial _{\mu} \phi)^2\over 2}+{m^2
 \phi^2 \over 2}+ m g_4 \phi^4 + g_6 \phi^6\big]. 
\label{eqn_Omega}
\ee
Note that quartic term in it is proportional to the $first$ power of the same
$m$ as is quadratic in the $\phi^2$ term. Indeed, 
at CP, when $m\rightarrow 0,\xi\rightarrow \infty$,  only the  $\phi^6$ term remains.

(A side comment: numerical simulations of Ref.\cite{hep-lat/9401034} are done for several lattices, but with constant ratio of the box size $L$ to
 the correlation length $\xi$, specifically $L/\xi=4.1$. This implies existence of about $(L/\xi)^3\sim 70$ statistically uncorrelated domains. Curiously, by numerical coincidence, a
 similar ratio (and number of domains) are expected 
 for fireballs corresponding to central heavy ion collisions and $\xi_{max}\sim 2\, fm$. Therefore,
 histograms  for mean field distributions  $P(\phi)\sim exp[-V_4\Omega(\phi)]$ from the paper are approximately
 the same as in these fireballs.)

Unlike in familiar 4 dimensions, in the $d=3$ setting of the Ising class we discuss the dimension of the field  is $[\phi]\sim L^{-1/2}$. Therefore all terms in (\ref{eqn_Omega})
scale as $L^{-3}$ for dimensionless couplings $g_4,g_6$.
In order to make field also dimensionless, let us 
define a scale $M$
by the beginning of near-$T_c$ scaling relation
with  critical index $\nu$ of the correlation length
\begin{equation}
m={1\over \xi} = M t^\nu  \label{eqn_xi_of_nu}
\end{equation} 
where we use standard dimensionless temperature variable $t\equiv (T/T_c-1)$. 
Let us also use it to define dimensionless field
\begin{equation}
\tilde \phi\equiv {\phi \over M^{1/2}}
\end{equation}
and rewrite potential for constant field as 
\begin{equation}
\Omega(\phi) =(V_3 M^3)
 \big[{t^{2\nu}	\tilde\phi^2 \over 2}
 +  g_4 t^{\nu} \tilde\phi^4 + g_6 \tilde\phi^6\big]
\end{equation}
The coupling values obtained by Tsypin  are
\begin{equation}
g_4=0.97 , \,\,\, g_6=2.05 \label{eqn_g4g6}
\end{equation}

We also compared these  lattice fits with 
exact RG solutions summarized in Ref. \cite{Berges:2000ew}.
From discussion in section 4.4 of that paper we extracted their polynomial fit, to 
\begin{equation} \label{eqn_derivativeU}
{\partial U\over \partial \phi}\sim (a_0 s+a_1 s^3+a_2 s^5+a_3 s^7)
\end{equation}
where $s$ is their re-scaled $\phi$. The fitted values are $$a_0 = 1.0084; \, a_1 = 3.1927; \,a_2 = 9.7076;\, a_3 = 0.5196$$ 
A drop from the six-field coefficient $a_2$ to the eight-field coefficient $a_3$ by a factor 20 confirms that truncation
of the eight-field term is indeed justified, as are
that for higher orders not used in the fit. Furthermore,
the values of other coefficients are 
in  a reasonably good agreement with (\ref{eqn_g4g6}).\\

The  scale $M$, defining the absolute size 
of the scaling window.
Below for estimates we will use $M= m_\sigma \approx 500$ MeV,  which implies that 
at the edge of this window, $ \xi(t=1)=0.4\,  fm$,
the $\phi$ exchange range is the same as in Walecka 
sigma meson exchange (which we subtract from the contribution of diagram (a) in forthcoming calculations of $\Delta F$).
So, with this choice we have zero effect at $t=1$
from (a) and negligible many-body forces.
It is  of course not universal, and it can be that scaling window is smaller, e.g. $M\sim 1 fm^{-1}$.

For orientation, with such choice of scale, the value $\xi=2 \, fm$ 
(comparable to the cluster sizes) corresponds to
   $t^\nu\approx 1/5$ or $t\approx 0.077$.
The calculations and
plots below, e.g. Fig. \ref{fig:expabc}, are done for $t$ ranging from 0.077 to 0.5.

\begin{figure}
 	\centering
 	\includegraphics[width=0.7\linewidth]{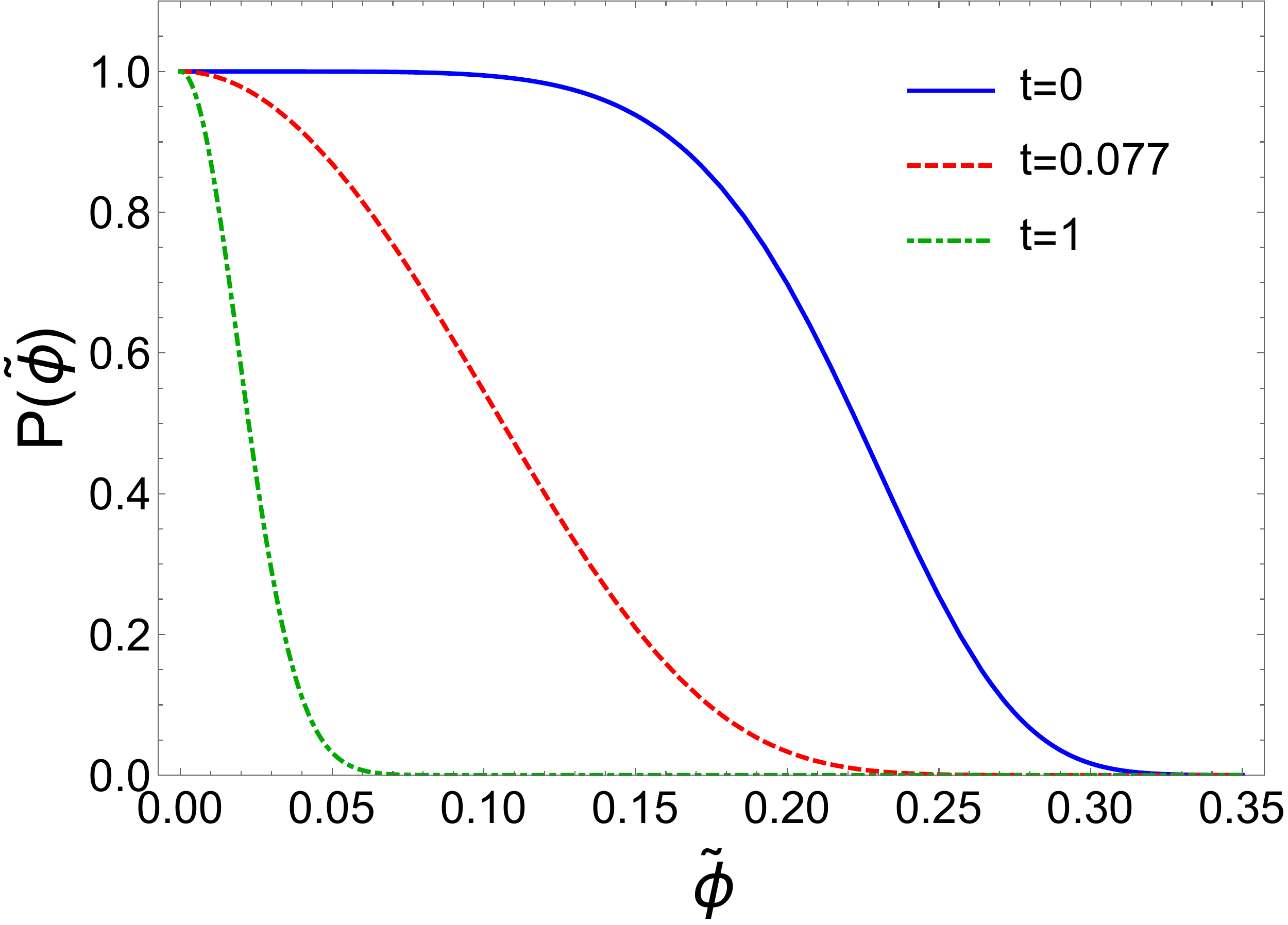}
 	\caption{
 (Color online) The universal probability distributions of dimensionless $\tilde \phi$ field. Far from CP corresponds to $t=1$ with Gaussian-like distribution, maximum expected correlation length $\xi = 2$ fm corresponds to $t=0.077$, and CP corresponds to $t=0$, here critical fluctuations are maximal and strongly non-Gaussian.    	
 }
 	\label{fig:fluctshapes}
 \end{figure}

The probability distribution depends on a prefactor of the scaled effective potential, the  3D volume  $V_3$ over which fluctuations are measured and $T$, in 
units of $M^3$ and $M$ respectively.
For  estimates 
one may take $V_3$ to be the volume of "preclusters" and use the kinetic freezeout temperature $T\approx 120$ MeV.  The resulting distribution is plotted in Fig. \ref{fig:fluctshapes}.

  \begin{figure}[h!]
 	\centering
 	\includegraphics[width=7cm]{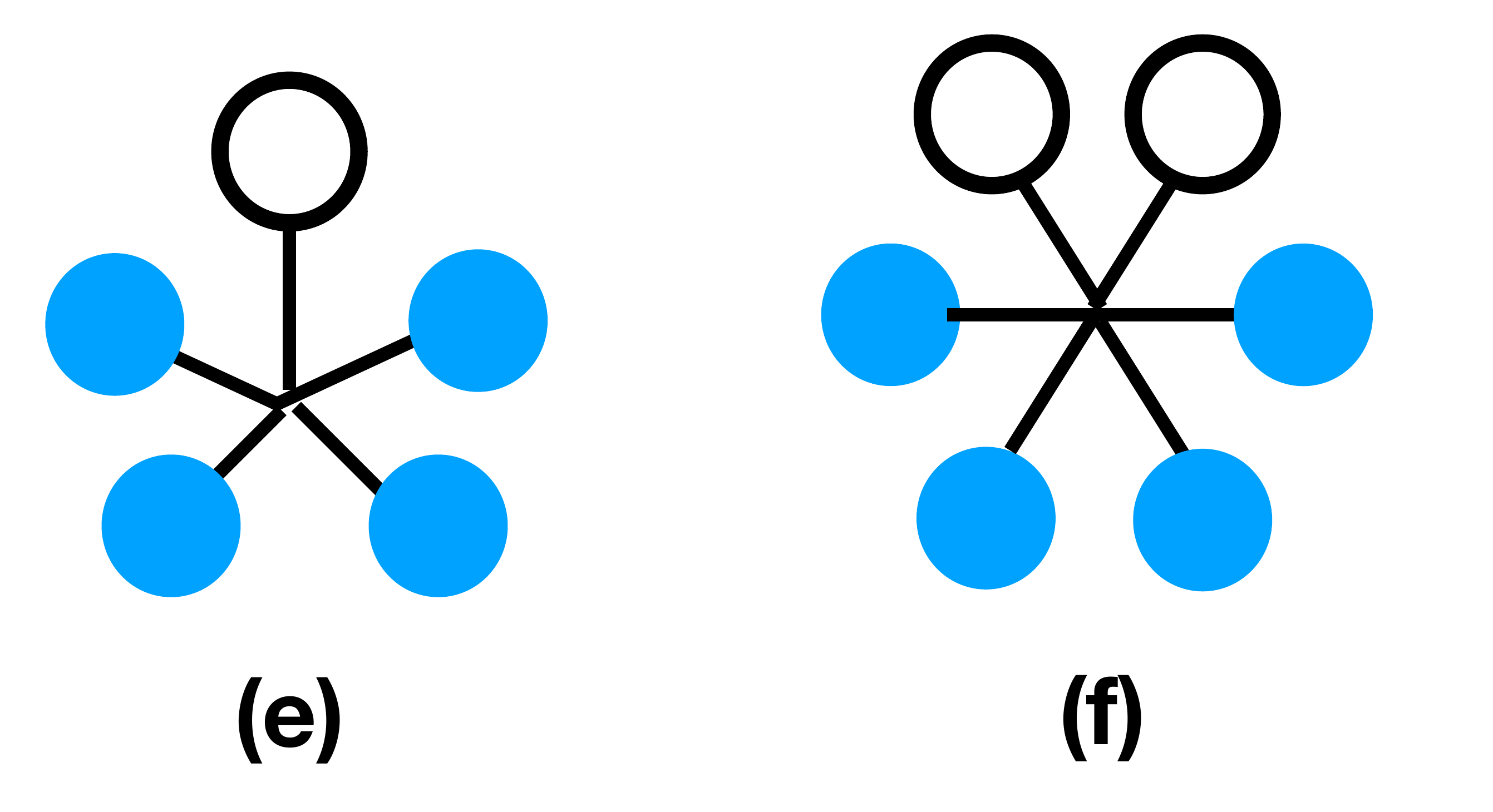}
 	\caption{Lowest order diagrams including five and six nucleons. Closed circles are those belonging to 4-N cluster, open circles indicate nucleons from the ``ambient matter" }
 	\label{fig:diagef}
 \end{figure}

 The assumed dominance of the 6-field coupling
 puts into question whether the original 4 diagrams 
 of Fig. \ref{fig_diag} would be enough, especially very close to the CP. Therefore 
 we introduce two more, shown in Fig. \ref{fig:diagef}.
 The diagram (f) for uncorrelated nucleons (4 in the cluster and 2 in ambient matter) can be estimated as
 \begin{equation}
 V_f/n\sim {g_c^6\over R_{cl} } g_6 \big( {\xi^{12} \over R_{cl}^{9} R_{amb}^3}\big) 
 \end{equation} 
 with two last brackets dimensionless. So, for $\xi\ll R_{cl} < R_{amb}$ 
it is extremely strongly suppressed, but if $\xi\sim R_{cl}\sim 2\, fm$ most of the suppression is gone.
It is this repulsive diagram alone which should be able to moderate
huge attraction due to diagram (a) at the CP.

\section{Deformed effective potential near the critical line} \label{sec_deformed}

The universal effective potential
discussed in the preceding section (\ref{eqn_Omega}) was defined $on$  the critical line.  Therefore it was symmetric under $\phi\rightarrow -\phi$  and  included only $even$ 
powers of $\phi$.
 However, in heavy ion collisions we expect 
 the endpoints of  evolution paths on the  phase diagram,
 known as
the {\em freezeout line},  to be located at  certain distance $below$ ( at lower $T$) 
 critical line. Such shift  modifies 
the effective potential. In particular,   the maximal value of the correlation length $\xi=1/m$ gets limited. Also the   $\phi\rightarrow -\phi$  symmetry is broken and
odd powers of $\phi$ appear. As we now detail, it turned out to be very important for the estimated  many-body forces.

We thought of two approaches to define the $deformed$ effective potential:\\
(1) One general way is to
start with the universal Equation of State (EOS) on the 2D plane of the Ising variables, the reduced $t$ and
the magnetization $M$, and then map it to QCD phase diagram. This approach, started in the epsilon-expansion
framework, was used by Nonaka and Asakawa \cite{nucl-th/0410078}, and Stephanov \cite{1104.1627}. We followed it to some
extent, and put some the related formulae and one plot in Appendix C.\\
(2) Another is to use the effective potential on the critical line, defined in the previous section,
and  calculate its deformation by a linear term $J \phi$,  assuming that $J$ 
{\em remains constant} at the freezeout  line. Using it, we calculate the deformation of 
effective potential shape and then use the coefficients of $\phi^3,\phi^4$  as effective nonlinear couplings
$\lambda_3,
\lambda_4$.

The first effect of the deformation by 
 $\tilde J \tilde \phi$ term is a
shift of the maximum  away  from the symmetry point $\phi=0$. Location of the new maximum $\tilde\phi_0(\tilde J)$  is to be found from solving polynomial equation
\begin{equation} {\partial \Omega _{def} \over \partial \tilde\phi}(\tilde\phi_0)=\tilde J
\end{equation} 
which, with our truncation, is of the 5th order.
As an example, for $\tilde J=1/100$ we
perform this procedure for various values of $t$.
In particular, the real roots of this equation are
$$ \tilde\phi_0(t=0.01)\approx 0.224,\,\,\, \tilde\phi_0(t=0.41)\approx 0.031
$$
We then 
rewrite the fluctuation field in the form 
\begin{equation} \tilde\phi=\tilde\phi_0 +
\delta
\end{equation} 

\begin{figure}[h]
	\centering
	\includegraphics[width=0.7\linewidth]{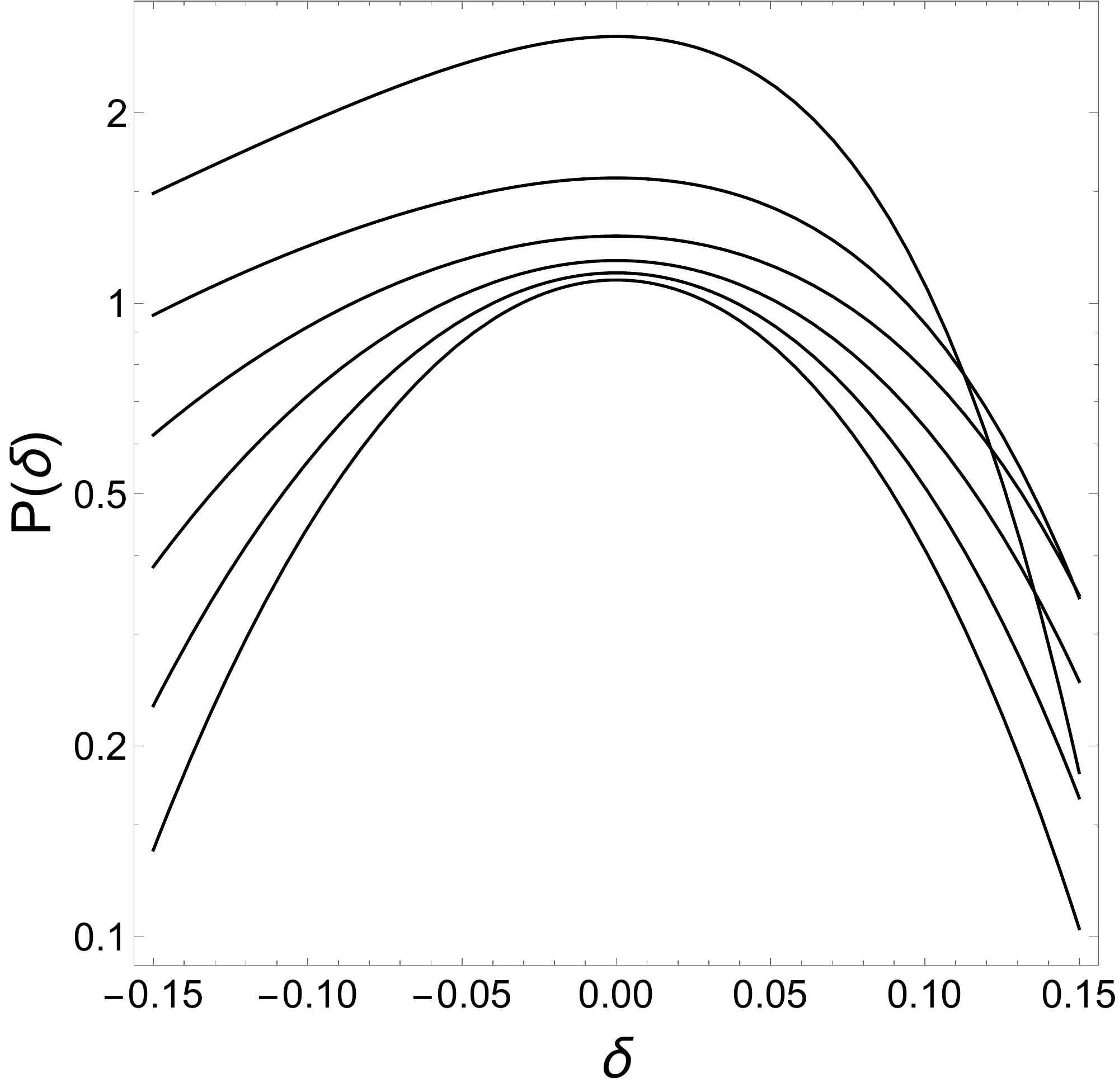}
	\caption{Probability distributions of the deformed action $P(\delta)$ distorted by $\tilde J=0.01$
		in the 4-nucleon cluster volume. The six curves, from top to bottom, correspond to values of $t=0.01,0.09,0.17,0.25,0.33,0.41$, respectively. The
		3D volume and mass prefactors are explained in the text.
	}
	\label{fig:expomega}
\end{figure}

and re-express the potential in terms of new fluctuation field $\delta$. This was done for all 
values of  $t$, for example the deformed potential
at $t=0.01$ takes the form
$$ \Omega_{def}(t=0.01)
\approx -0.0017 + 0.095 \delta^2 + 0.51 \delta^3 + 
1.60 \delta^4 + 2.75 \delta^5 + 2.05 \delta^6
$$
 Note that there is
no linear terms, but  other odd powers of $\delta$ are present. 

In order to get an idea about actual distributions of the fluctuating critical field one has to return to dimensionful prefactor of the universal action, and also
select the scale at which the fluctuations will be studied. The probability distribution of homogeneous 
fields is given in Eq. (\ref{eq:def}) where $V_3$ is the 3D volume,
made dimensionless by the 3rd power of basic scale $M$. Using the volume of the cluster  $V_3=(4.3*10^4$ fm$^9)^{1/3}$, one finds a very large
product of the first bracket, $\sim 550$. Yet since
small $\delta$ appears in high powers, one gets the distributions shown in Fig. \ref{fig:expomega}.
 While it is approximately Gaussian for larger $t$ (bottom curves), it becomes quite strongly 
deformed close to CP.

\begin{figure}[h!]
	\centering
	\includegraphics[width=0.7\linewidth]{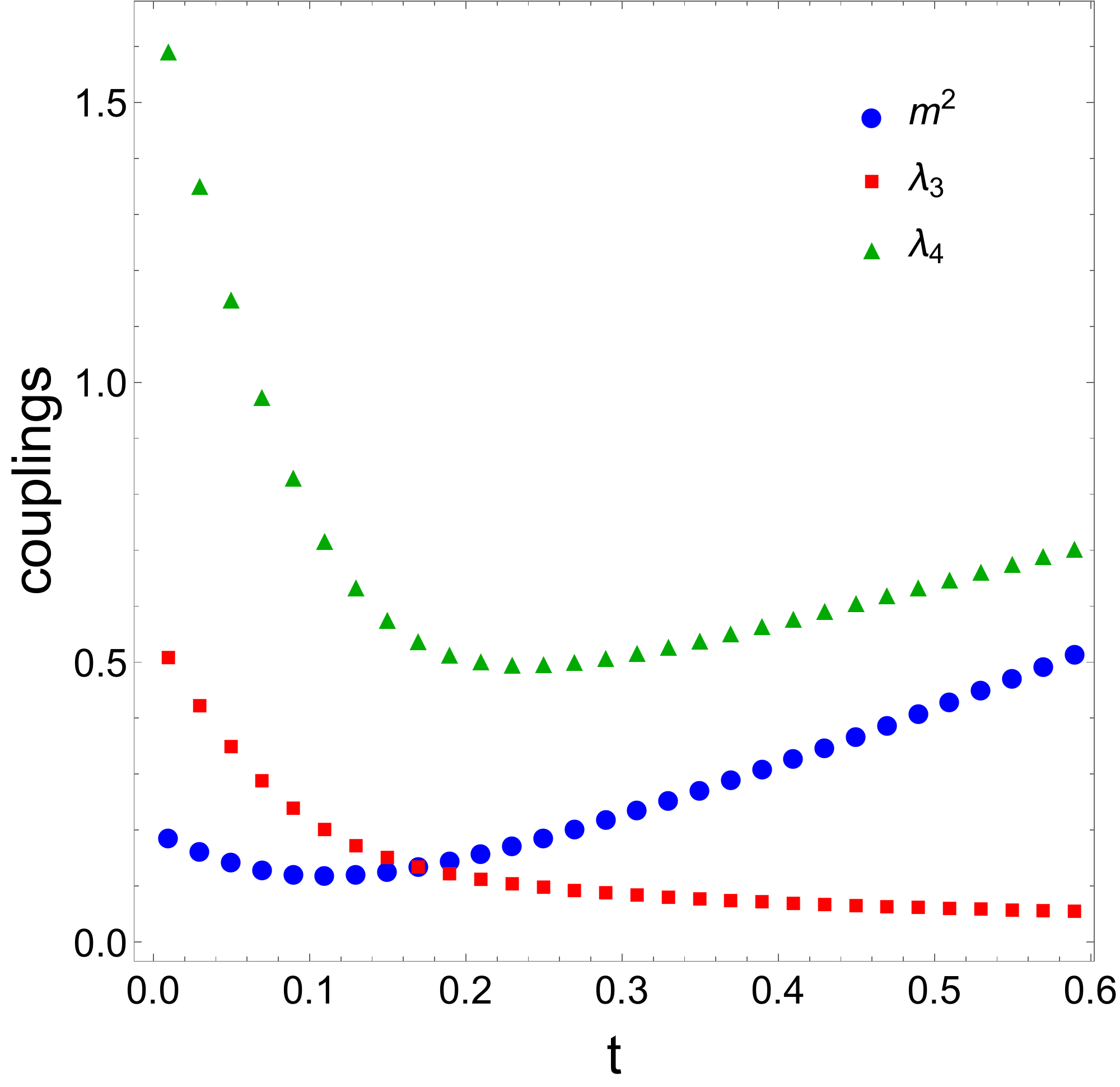}
	\caption{(Color online) The dependence of $m^2=1/\xi^2$, and the triple and quartic couplings,
		for the  effective action deformed by $J=1/100$, on scaled temperature $t$. }
	\label{fig:deformedj01}
\end{figure}

The dependence of $m^2=1/\xi^2$, and the triple and quartic couplings
from the deformed effective action on $t$ is shown in Fig. \ref{fig:deformedj01}. Note significant growth 
of the coupling near CP (left). Note also that at small $t$ the inverse correlation length $m$ does not go to zero, although
it remains small.

\begin{figure}[h!]
	\centering
	\includegraphics[width=1\linewidth]{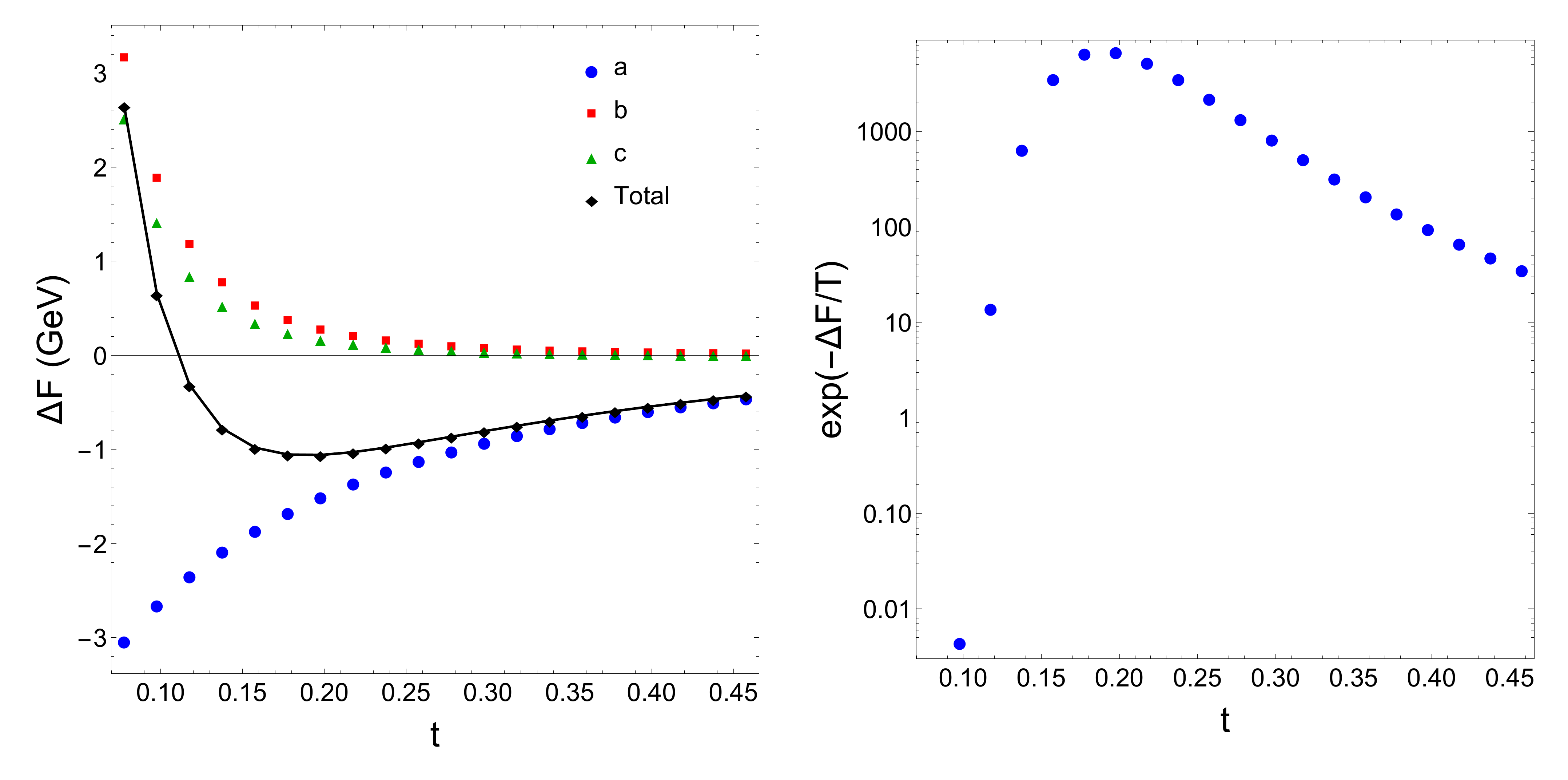}
	\caption{(Color online) Left plot: contributions to the change in free energy $\Delta F$ of a 4-nucleon cluster of size $\rho=2$ fm due to individual diagrams (a,b,c) and their total combined contribution, with coupling $g_c^2/4\pi = 10$, versus $t = T/T_c - 1$. Diagram (a) has had the unmodified binary interaction subtracted out as described in the text. Right plot: Boltzmann factor of the change in potential $\exp(-\Delta F/T)$ with $T= 120$ MeV, versus t. Note that the leftmost point, $t=0.077$ (not shown) has Boltzmann factor $\simeq 10^{-10} $.	
	}
	\label{fig:expabc}
\end{figure}

\begin{equation}
P(\delta) \sim exp\big[- (V_3 M^3)\Omega_{def}(\delta)  \big]
\label{eq:def}
\end{equation}

Unfortunately, the real fluctuating fields are not homogeneous, and so these distributions serve only for orientation. What one needs to do is to evaluate 
the diagrams with propagators containing appropriate
correlation length for each $t$. The 
value of the nonlinear couplings $\lambda_3,\lambda_4$
should be taken as coefficients of $\delta^3,\delta^4$ terms. In the case of $\lambda_3$ a factor of $M$ is inserted to restore it to its dimensionful form.

The free energy density of a cluster divided by the nucleon density we define for binary term as follows
\begin{equation}
F_a=-{4 \cdot 3\over 2} {g_c^2 \over 4\pi r} exp\big[-rMt^\nu\big], 
\end{equation}
	for three-body force as
\begin{equation}
F_b=4 \cdot 3! \lambda_3 M \big({g_c \over 4\pi}\big)^3 V_b(\rho M t^\nu), 
\end{equation}
	and for (diagram c) four-body force as 
\begin{equation}
F_c=4! \lambda_4 \big({g_c \over 4\pi}\big)^4 V_c(\rho M t^\nu). 
\end{equation}	
Here, inter-nucleon distance $r$ and hyperdistance $\rho$ are related as they are in the tetrahedral cluster, $r= \sqrt{2/3} \rho$.

Our task is now to combine all terms and see how they affect the 4-body clusters. 
In Fig. \ref{fig:expabc} (left) we show the results of our calculation of the free energies at  ten values of $t$, increasing from $t_{min}=0.077$ and
 corresponding to diagrams (a,b,c), separately and in sum.
A very large attractive contribution (calculated in some earlier works)
is in fact compensated by 3- and 4-body repulsive terms,
so that the sum becomes positive for $t<0.11$, before
the maximal correlation length is reached.

The contribution of diagram (d) 
\begin{equation}
	F_d=-{4! \over 8\pi}\lambda_3^2  M^2 \big({g_c \over 4\pi}\big)^4 V_d(\rho M t^\nu) 
\end{equation}
is not included in the plot because it turned out to be 
small, well inside the uncertainties. In particular,
its largest value (at $t_{min}$, the leftmost point in Fig. \ref{fig:expabc} (left)) is only $-164$ MeV. 

Since, in the left plot of Fig. \ref{fig:expabc}, it is hard to read the magnitude of
the attractive effect on the r.h.s. , we separately show how this free energy translates into the
probability of precluster production, $\exp(-\Delta F/T)$ in the right plot. In it one finds that attractive force is strong enough to enhance clustering, by a few orders of magnitude at distance $t=0.2$
from the CP. At the same time it plunges well below 1 due to repulsive many-body forces at smaller $t$ (closer to the CP). This is  the ``non-monotonous signal"  
we speak about.

 Let us remind that very strong effects displayed  in Fig. \ref{fig:expabc} were shown as a function of $t$,
 on a line close to the critical line distorted by $\tilde J=1/100$, for clusters of fixed size  $\rho=2\,fm$. We selected this size as characteristic
 of pre-clusters as PIMC calculation with conventional
 nuclear forces.
 
\begin{figure}
 	\centering
 	\includegraphics[width=1.0\linewidth]{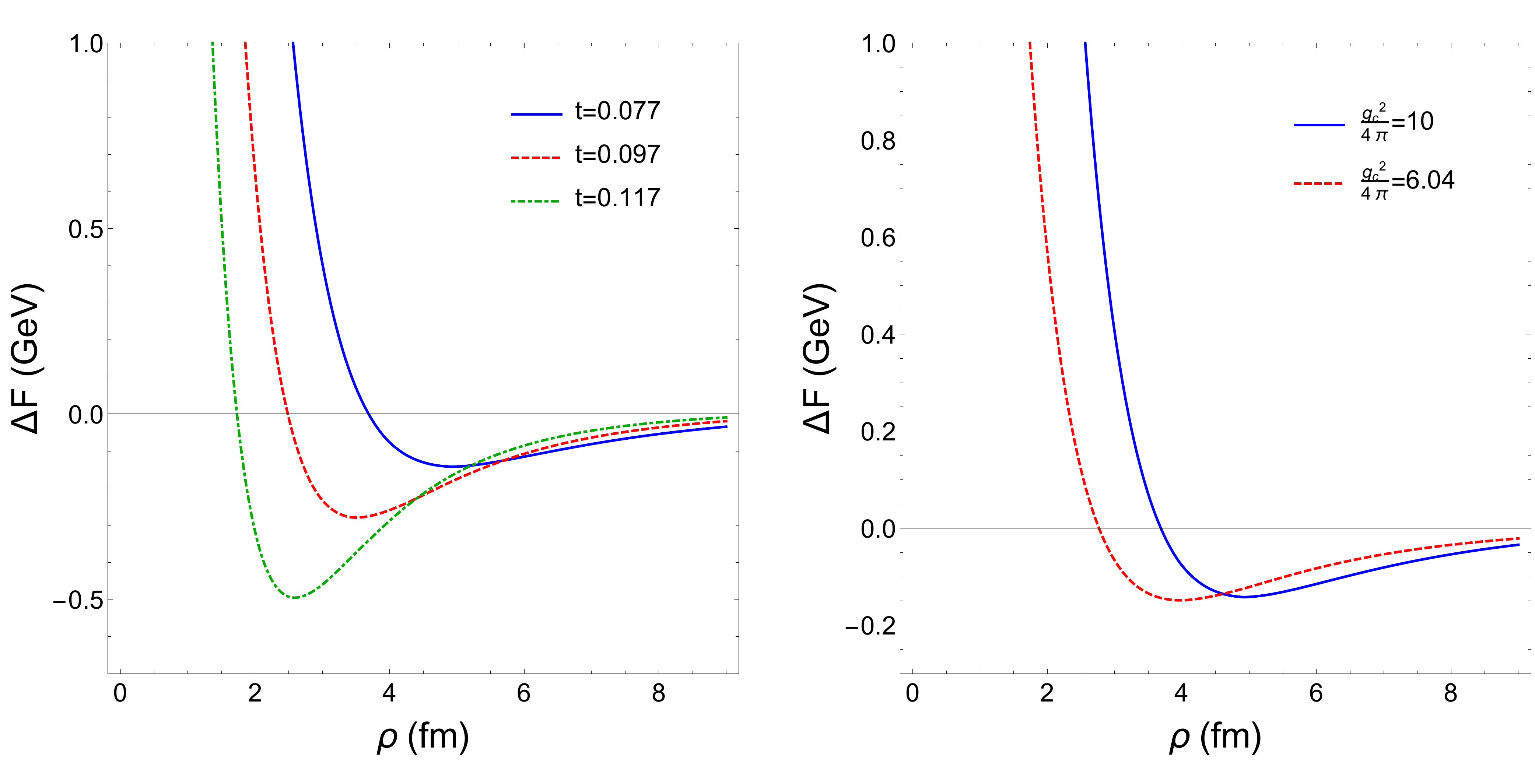}
 	\caption{(Color online) Left plot: The change in effective potential $\Delta F$ as a function of cluster size $\rho$ with $g_c^2/4\pi = 10$ for three values of scaled temperature $t$. Right plot: The change in effective potential $\Delta F$ as a function of cluster size $\rho$ at $t=0.077$ for two values of the nucleon-critical mode coupling $g_c$. In both plots, diagram (a) has had the unmodified binary interaction subtracted out as described in the text.}
 	\label{fig:vofrho}
 \end{figure}

  Another perspective on the problem is obtained if one fixes
 $t$, say to values rather close to CP, just above 
 $t=0.077$ with correlation lengths just below $\xi = 2$ fm, and  plot the total energy of the cluster as a function of its size $\rho$, see
  Fig. \ref{fig:vofrho}. One can see from it that
  while for $\rho<2$ fm the potential is indeed repulsive and
  much larger than $T\sim 100$ MeV, it is very rapidly
  changing for larger sizes. As one approaches CP, the size of this repulsive region increases and the maximum depth of the attraction decreases. In particular, near the minimum at $\rho\approx 4$ fm, $-\Delta F/T\approx 2$ at the smallest value of $t$.
  Therefore, here instead of suppression one finds enhancement in the production of clusters of a larger size relative to PIMC is by factor $\exp(2)\sim 7$, rather than by three orders of magnitude, as in Fig. \ref{fig:expabc}. This qualitative behavior remains unchanged for different reasonable choices of the nucleon-critical mode coupling. Varying this coupling modifies the size of the repulsive region, while keeping the maximum depth of the attraction relatively fixed, as seen in Fig. \ref{fig:vofrho} (right). 
  
  It might
  be tempting to conclude that accounting for many-body forces simply modifies clusters to be of the size $\rho > 3$ fm rather than $\sim 2$ fm as was seen in PIMC calculations. Such a conclusion would however
  be rather meaningless, since at such size the effective cluster density would not be any different from that of ambient matter. In other words,
  there would be enhancement, but feed-down from such 
  large clusters to light nuclei production would be negligible, as the clusters are much larger than the excited $^4He$ states which feed down into light nuclei.

 Concluding our calculations, we remind the reader that while
  in this paper we focused only on 4-body clusters,
  there are of course larger ones. For them
 one should also include five and six-body forces. 
Note that the deformed effective action  predicts them to be also repulsive,
and even larger . Therefore,
our main finding -- suppression of all forms of clustering in the vicinity of the CP -- should hold, even if all possible clusters are included.

 \section{Summary, discussion and experimental observables}\label{sec_discussion}
 Let us start by reminding the reader the paradox (pointed out in Ref.  \cite{Shuryak:2006eb}):
the effect of binary forces induced by long-range critical mode {\em at  CP}, with $\xi\rightarrow \infty$, is
 catastrophic. Indeed, 
   if all $N(N-1)/2\sim 10^4$ pairs of nucleons in the fireball be attracted to each other by
 Newton-like  potential, the fireball would implode, like in a gravitational collapse.
 
 The resolution of this paradox is one of the main 
conclusions of this paper. Large correlation length $\xi$  
 generates also  {\em repulsive many-body forces},  
 strong enough  to mitigate the binary attraction
 and  reverse the trend, 
 $suppressing$ preclustering close to CP. 
 
With this qualitative conclusion, let us 
discuss the uncertainties involved. Many features of the CP are known, as it is supposed to belong to the 3D Ising universality class. Yet some basic mass scale $M$ 
and the critical mode coupling $g_c$ are non-universal and remain unknown. 
 Changing $g_c$ will modify
the overall scale of the predicted effects, as $N$-body interactions depend on $g_c^N$. The mass scale $M$ appears directly in the 3-body term and affects the mapping between $t$ and $\xi$. While their values are not known, we have used physically-motivated estimates -- the values should be comparable to the nucleon-sigma coupling $g_{\sigma}$ and the sigma mass $m_{\sigma}$, respectively. Additionally, the external current $\tilde{J}$ deforms the potential.  At present, we have chosen $\tilde{J}$ to be small  to reflect the closeness of the critical and freezeout lines. 
Fortunately, dependence on the specific value of $J$ is weak, e.g. $\langle \phi \rangle \sim J^{1/5}$.
 Needless to say,
all such non-universal parameters may be fitted
to the data, once the CP is found.

\begin{figure}[h!]
	\centering
	\includegraphics[width=0.7\linewidth]{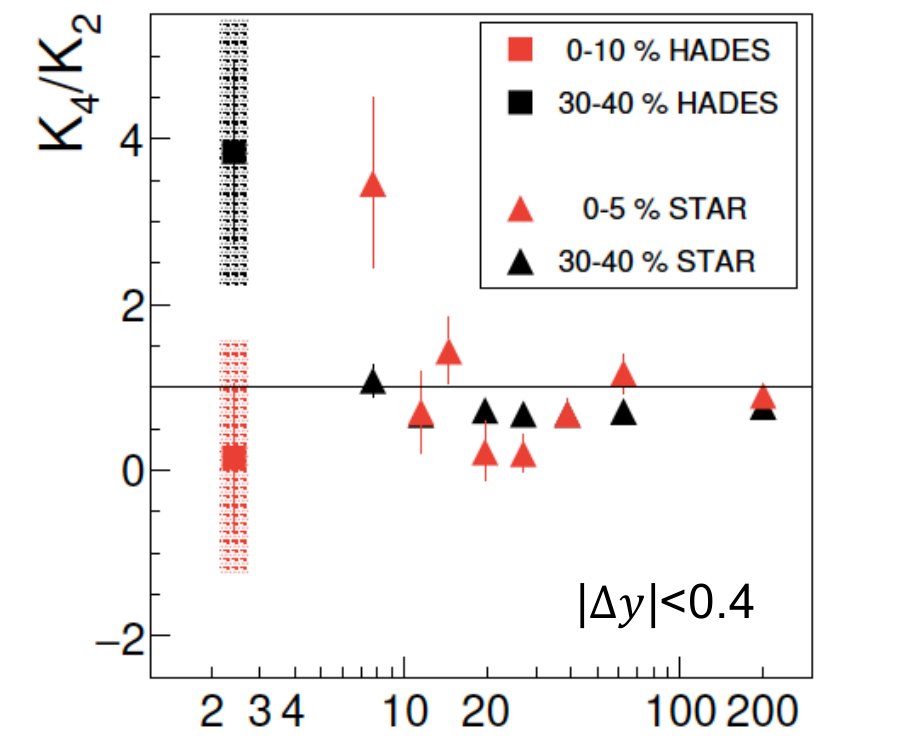}
	\includegraphics[width=0.7\linewidth]{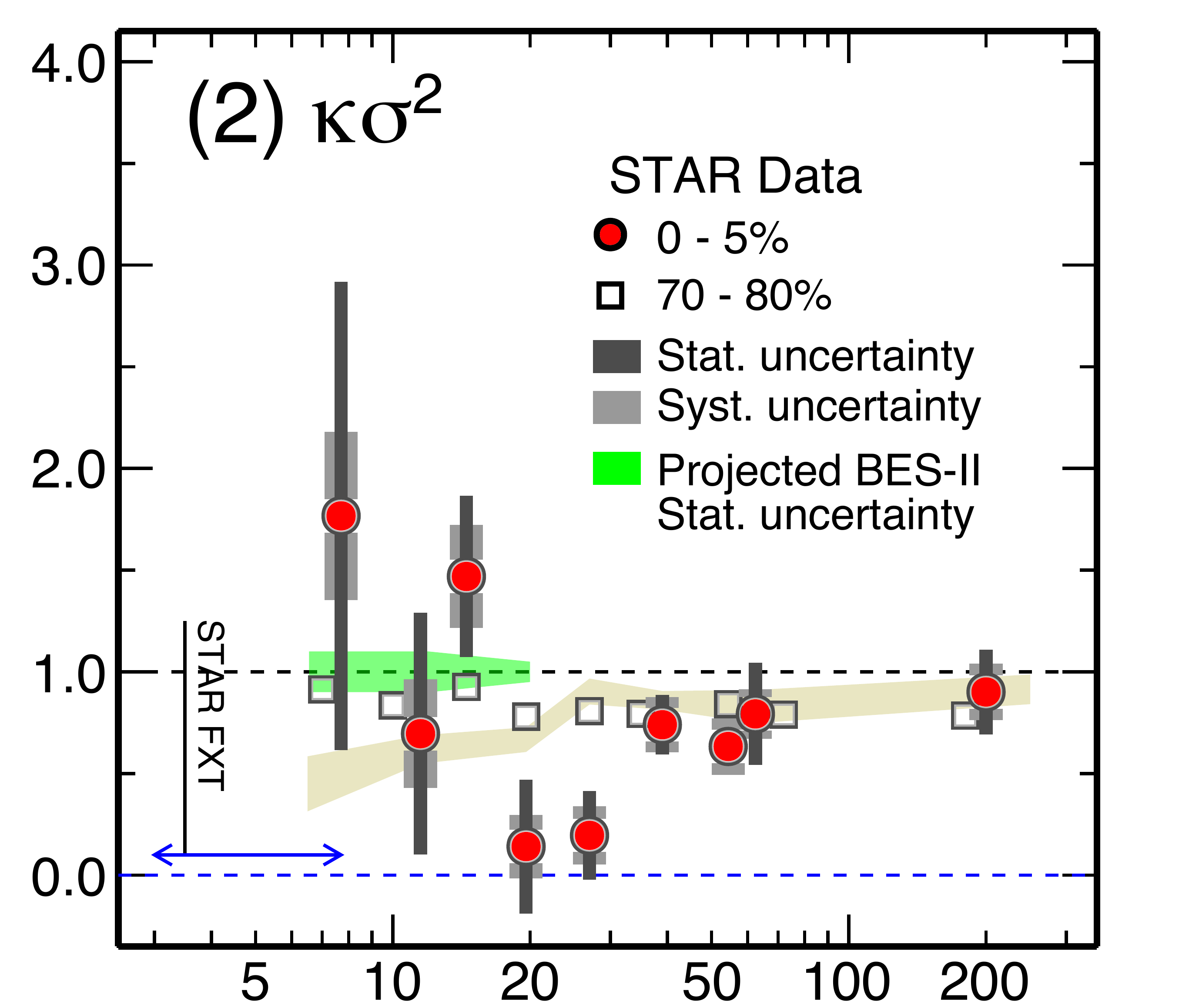}
\caption{(Color online) Upper plot: The kurtosis ratio $K_4/K_2$ from STAR and Hades experiments, versus $\sqrt{s}$, from \cite{Adamczewski-Musch:2020slf}.
		Red points show the most central bins,  black points for mid-central one, $30-40\%$.\\
Lower plot: more recent STAR results, corrected in \cite{Adam:2020unf}.  	
	} \label{fig:kurtosisexp}
\end{figure}

 Let us now proceed to the status of experimental
 observables.  The summary of  kurtosis data of the net proton distribution is
  shown in Fig. \ref{fig:kurtosisexp}. The upper plot is
  earlier summary, the lower one is from recent STAR publication \cite{Adam:2020unf}.  More strict event selection applied have basically modified one point,
  the central bin of $7.7 \, GeV$ run.  
  
  Interpreting summary plots one should keep in mind  that  detectors involved in these and next plots -- STAR at  BNL RHIC, HADES at GSI ,
  NA49 at CERN SPS and ALICE at LHC --
 have completely different kinematic settings,  acceptances and use different extrapolation procedures.
 Therefore,  comparison of their points needs to be done  with care. As one can see, the errors are still large. The lower plot indicate projected accuracy
 of BES-II program (green shaded area near 1).

  On the other hand, {\em inside each group} the centrality bins are supposed to be processed in exactly the same way. If one trusts the centrality dependence of each set,  one finds a striking $reversal$, between
 the
  STAR centrality dependence at $7.7\, GeV$ and that reported by HADES at $2.4\, GeV$. 
  Also both plots show clear depletion near $20\, GeV$,
  in central relative to peripheral bins.

 \begin{figure}[h!]
 	\includegraphics[width=0.5\linewidth]{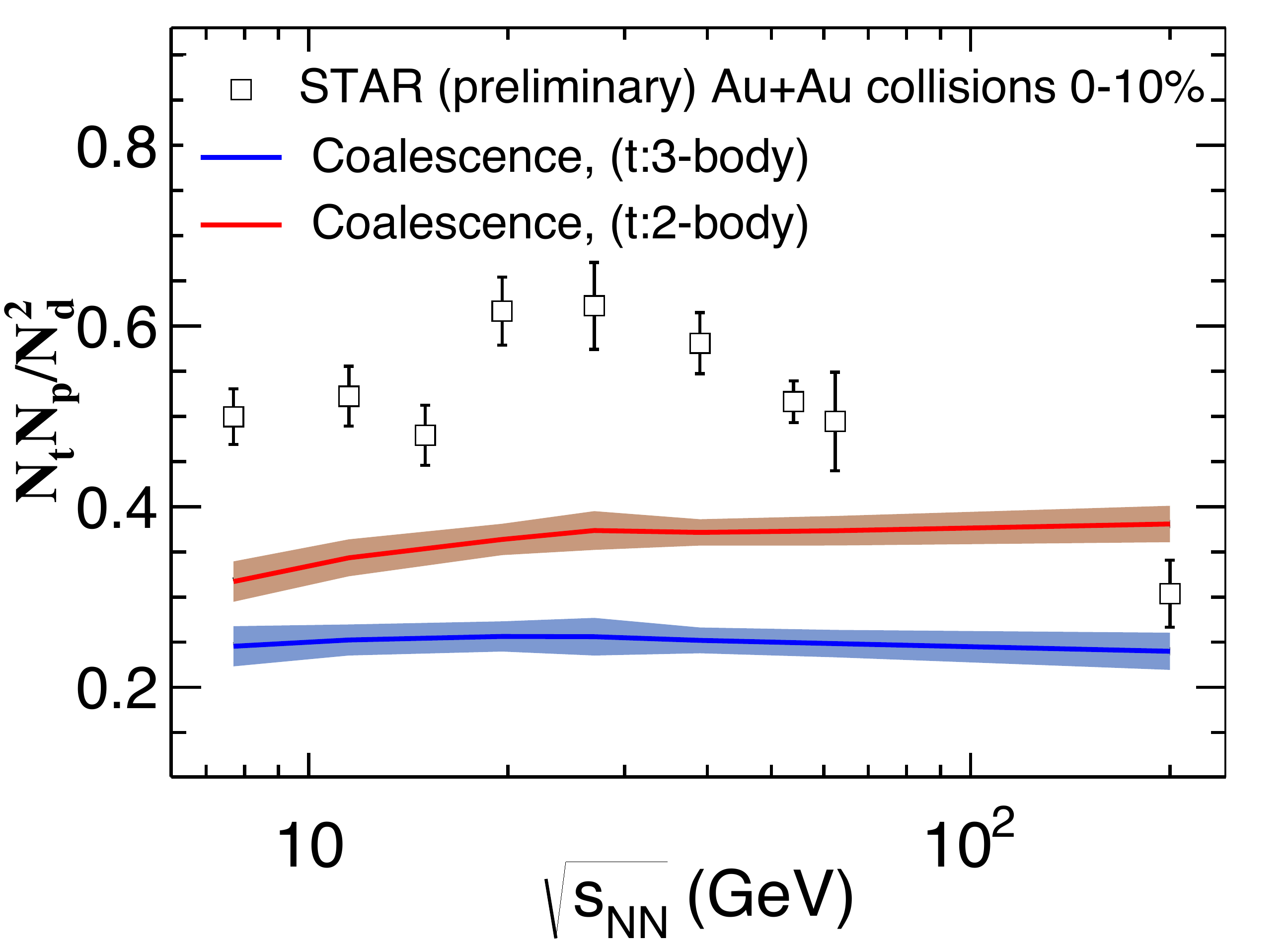}
 	\includegraphics[width=0.4\linewidth]{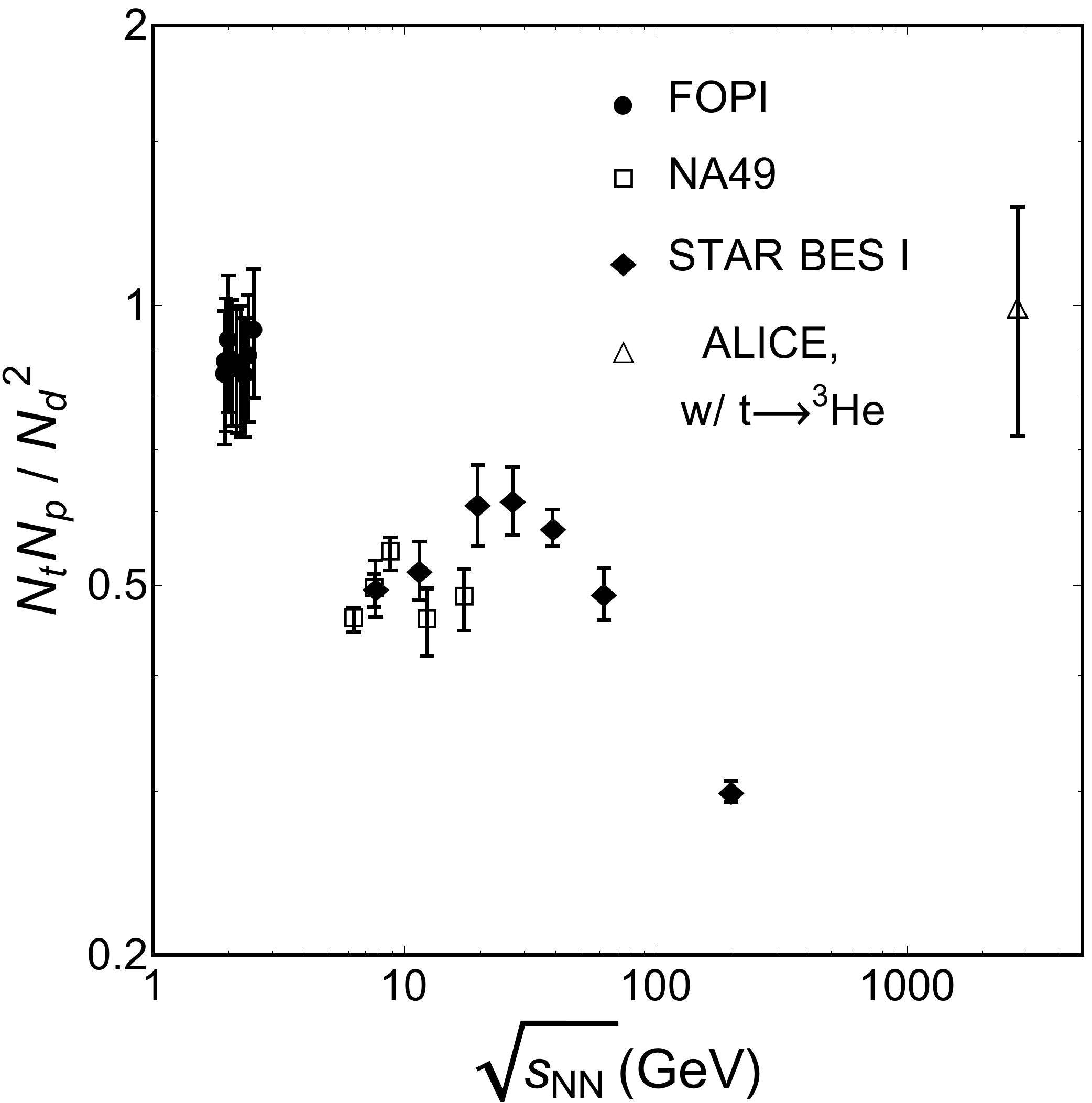}
 	\caption{(Color online) Left plot: The ratio of yields of tritium,  protons  normalized to deuterium $t\cdot p/d^2$, from a cascade code  \protect\cite{Zhao:2020irc}, points are from STAR collaboration.
 		Right plot: Compilation of experimental  data for the same ratio  from \cite{Shuryak:2020yrs}.  Note that it has different range from the left plot, and is a log-log plot.
 	} \label{fig:tritium}
 \end{figure}

   Another observable sensitive to preclustering is additional  feed-down into
 production of light nuclei. The most sensitive to four-nucleon clusters are  $t$ and $^3He$. The compilation of the data for the ratio of yields $N_t N_p/N_d^2$ from \cite{Shuryak:2020yrs} is shown in Fig. \ref{fig:tritium}.
 
 This ratio is selected because in it the main driver 
 of light nuclei yields -- the factors of fugacity $exp(\mu/T)$ -- cancel out. The left plot, from \cite{Zhao:2020irc}, is theoretical  predictions resulting from a 
 state-of-the-art cascade code. It reproduces many features of heavy ion collisions but does not have
 preclusters and feed-down into tritium. As one can see, it  basically no
 dependence of the ratio on collision energy is predicted. Furthermore, these predictions are well below STAR data points, except for
 the rightest point, at $\sqrt{s}=200\, GeV$, which is consistent with simple ratio of number of states for these nuclei, equal to 0.29.  
  The right plot is larger data compillation, including all available data
 from   experiments indicated on the figure.

  So, experimental data on both kurtosis and light nuclei ratio  show some 
  hints for non-monotonous patterns of the type we discussed. 
 Let us enumerate them once again:

\begin{enumerate} 
\item  The most dramatic change in Fig. \ref{fig:kurtosisexp}(upper) is the reversal of centrality dependence between  $\sqrt{s}=7.7\, GeV$ and $2.4\, GeV$ already noticed.  Yet the error bars are large.
 
\item There seems to be 
 smaller  dip in kurtosis at $\sqrt{s}\sim 20\, GeV$ -- two red triangles corresponding to central collisions. Combing
  the errors of those, one sees that deviation from the default value of 1 should indicate some real effect rather than mere statistical
  fluctuation.  

\item Fig. \ref{fig:tritium} (lower) indicate a dip
between HADES data on the left and the lowest energy at
STAR and NA49. 

\item There are also (admittedly weaker) indications of
another minimum, again at $\sqrt{s}\sim 20\, GeV$
 
\end{enumerate}

Finally,
 as
it has been pointed out
 in \cite{Shuryak:2019ikv}, the  
 preclusters  decays  have certain binary modes, e.g. 
 $p+t$ and $d+d$, potentially a tool to  
 monitor preclustering and feed-downs directly. 
 High statistics of
 BES-II data should allow a dedicated search for them.

 \begin{acknowledgments}
 	
 	This work was supported in part by the U.S. Department of Energy, Office of Science, under Contract No. DE-FG-88ER40388. We would like to thank Juan Torres-Rincon for multiple helpful
discussions.
 	
 \end{acknowledgments}

\appendix
\section{Jacobi coordinates and hyperdistance  for 4-nucleon cluster} 
The first standard step in many-body physics is the separation of the center of mass motion from the
relative coordinates. It is usually done using Jacobi coordinates, which for the $A=4$ case are
\begin{align} 
\vec \xi_1 & =\frac{\vec x_1 - \vec x_2}{\sqrt{2}}, \,\,\, \vec \xi_2=
\frac{\vec x_1 + \vec x_2 - 2 \vec x_3}{\sqrt{6}} \ , \\ 
\vec \xi_3 & = \frac{ \vec x_1 + \vec x_2 +\vec x_3 - 3 \vec x_4}{2 \sqrt{3}} \ . 
\end{align}

The radial coordinate, or hyperdistance, is defined as
\be \rho^2=\sum_{m=1}^3  (\vec \xi_m)^2=\frac{1}{4} \left[ \sum_{i\neq j} (  \vec x_i-\vec x_j)^2 \right] \ .
\label{rho_definition}
\ee
The radial part of the Laplacian in these Jacobi coordinates is $\psi''(\rho)+8\psi'(\rho)/\rho$,
and using the substitution 
\be \chi(\rho)=\psi(\rho) \rho^4 \label{substitution} \ , \ee
one arrives to the conventional-looking Schr\"odinger equation for $K=0$ harmonics
\be 
\frac{d^2 \chi}{d\rho^2} -\frac{12}{\rho^2} \chi-\frac{2m_N}{\hbar^2} [W(\rho)+V_C(\rho)-E] \chi=0 \ ,
\label{eqn_radial_for4}
\ee
where $W$ is the projection of the potential to this harmonic.

\section{Scaling exponents in Ising universality class} 
The main variables are $t=(T-T_c)/T_c$ and ``magnetization" (for $t<0$) $M$.
The critical phase diagram in $t,M$ has critical point at its origin.
Although those are well known, we remind the reader the definitions and values of the values of the scaling exponents
in Ising universality class.
The magnetization scales as
\be M\sim (-t)^\beta ,\,\,\,\,\, \beta\approx 0.326. \ee

The correlation length scales as
\be \xi\sim t^{-\nu},\,\,\,\,\, \nu=0.6299 .
\label{eqn_nu}
\ee

\section{The dependence of the correlation length on $t$, in epsilon expansion method and constant magnetization}
Thermodynamics near the critical point, and the correlation length were discussed by Nonaka and Asakawa \cite{nucl-th/0410078},
based on results of epsilon expansion to order $\epsilon^2$ in Ref. \cite{Zinn-Justin}.
The correlation length squared has the form
\begin{equation}  \label{eqn_xi2_from_eps}
\xi^2 = \xi_0^2M^{-2 \nu\beta} g\big({| t |\over |M|^{1/\beta} }\big), \end{equation}
where 
\ba  \label{eqn_g_of_x}
g(x)=
6^{-2\nu} z
  \Big[
   1 -  {\epsilon \over 36}((5 + 6 log(3)) z - 6(1 + z)log(z)) + 
    \epsilon^2 \big[{1 + 2*z^2 \over 72} log(z)^2 + \\
   {log(z)  z (z - 1/2) (1 - log(3)) \over 18} -    
       (16 z^2 - 47 z/3 - 56/3){log(z) \over 216} +\nonumber \\
    (101/6 + 2/3*Int + 6*log(3)^2 + 4*log(3) - 10)*{z^2 \over 216} -  \nonumber \\
    (6*log(3)^2 + 44*log(3)/3 + 137/9 + 8*Int/3)*{z \over 216} \big]  \nonumber
  \Big]
   \ea

where  $\epsilon=4-d$, 1 in $d=3$ case,  shown for consistency with its derivation. Here the argument of the r.h.s. is
$z\equiv 2/(1+x) $ and $$Int= \int_0^1 {ln[x(1-x)] \over 1-x(1-x) }dx\approx -2.344$$.

Its usage is made in the next section.

\begin{figure}[htbp]
\begin{center}
\includegraphics[width=8cm]{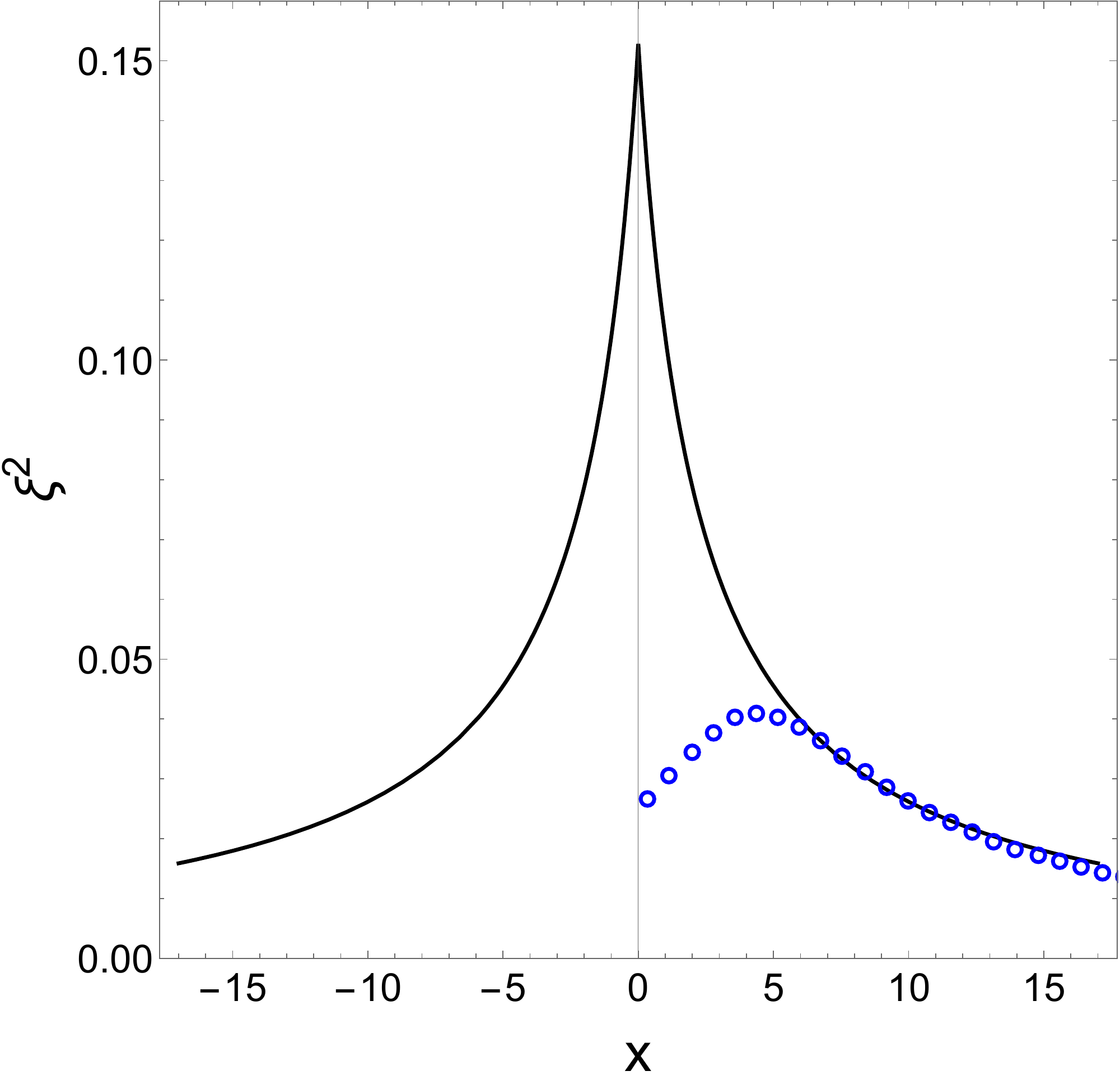}
\caption{(Color online) The line 
	corresponds to $\xi^2$ from
	Eq. (\ref{eqn_xi2_from_eps}) as a function of $x$ in the assumption of constant magnetization. Open points correspond to
assumption used in the main text, of constant external current $\tilde{J}=1/100$.   }
\label{fig_g_of_x}
\end{center}
\end{figure}

 \section{Alternative treatments of the effective potential away from the critical line}
 As we have seen above, the interrelation between the binary and many-body forces strongly depends on the magnitude
of the correlation length $\xi$, which enters in large and different powers in the different diagrams.
Mapping of the Ising variables to QCD phase diagram 
is a nontrivial problem,
discussed in Refs. \cite{1104.1627,nucl-th/0410078}.

 In the main text we followed a simplified procedure to calculate
the deformed potential, assuming certain $constant$ ($t$-independent) value of the external current $J$. It included calculation of the dependence of the magnetization (called there $\phi_{0}$) on $t$.

Here we would like to compare it to another simple 
map,  assuming instead $\phi_{0}=M=const(t)$ and using epsilon expansion expression (\ref{eqn_xi2_from_eps}).
The complicated function  is used  as a function of variable 
$$x={ | t |\over |M|^{1/\beta} }$$
 The  correlation length squared  following from it 
is shown in Fig. \ref{fig_g_of_x} by a line,
 compared
to the coefficient of $\delta^2$ term in those potentials, presented by open points. It shows
that while two alternative assumptions agree at larger $t$, they produce very different values of the correlation length very close to the critical point. 
Deformation of the potential by $\tilde{J}=1/100$ imposes a stronger limit on the correlation length.

\bibliography{clusters.bib}

\begin{thebibliography}{17}%
\makeatletter
\providecommand \@ifxundefined [1]{%
 \@ifx{#1\undefined}
}%
\providecommand \@ifnum [1]{%
 \ifnum #1\expandafter \@firstoftwo
 \else \expandafter \@secondoftwo
 \fi
}%
\providecommand \@ifx [1]{%
 \ifx #1\expandafter \@firstoftwo
 \else \expandafter \@secondoftwo
 \fi
}%
\providecommand \natexlab [1]{#1}%
\providecommand \enquote  [1]{``#1''}%
\providecommand \bibnamefont  [1]{#1}%
\providecommand \bibfnamefont [1]{#1}%
\providecommand \citenamefont [1]{#1}%
\providecommand \href@noop [0]{\@secondoftwo}%
\providecommand \href [0]{\begingroup \@sanitize@url \@href}%
\providecommand \@href[1]{\@@startlink{#1}\@@href}%
\providecommand \@@href[1]{\endgroup#1\@@endlink}%
\providecommand \@sanitize@url [0]{\catcode `\\12\catcode `\$12\catcode
  `\&12\catcode `\#12\catcode `\^12\catcode `\_12\catcode `\%12\relax}%
\providecommand \@@startlink[1]{}%
\providecommand \@@endlink[0]{}%
\providecommand \url  [0]{\begingroup\@sanitize@url \@url }%
\providecommand \@url [1]{\endgroup\@href {#1}{\urlprefix }}%
\providecommand \urlprefix  [0]{URL }%
\providecommand \Eprint [0]{\href }%
\providecommand \doibase [0]{http://dx.doi.org/}%
\providecommand \selectlanguage [0]{\@gobble}%
\providecommand \bibinfo  [0]{\@secondoftwo}%
\providecommand \bibfield  [0]{\@secondoftwo}%
\providecommand \translation [1]{[#1]}%
\providecommand \BibitemOpen [0]{}%
\providecommand \bibitemStop [0]{}%
\providecommand \bibitemNoStop [0]{.\EOS\space}%
\providecommand \EOS [0]{\spacefactor3000\relax}%
\providecommand \BibitemShut  [1]{\csname bibitem#1\endcsname}%
\let\auto@bib@innerbib\@empty
\bibitem [{\citenamefont {Stephanov}\ \emph {et~al.}(1998)\citenamefont
  {Stephanov}, \citenamefont {Rajagopal},\ and\ \citenamefont
  {Shuryak}}]{Stephanov:1998dy}%
  \BibitemOpen
  \bibfield  {author} {\bibinfo {author} {\bibfnamefont {M.~A.}\ \bibnamefont
  {Stephanov}}, \bibinfo {author} {\bibfnamefont {K.}~\bibnamefont
  {Rajagopal}}, \ and\ \bibinfo {author} {\bibfnamefont {E.~V.}\ \bibnamefont
  {Shuryak}},\ }\href {\doibase 10.1103/PhysRevLett.81.4816} {\bibfield
  {journal} {\bibinfo  {journal} {Phys. Rev. Lett.}\ }\textbf {\bibinfo
  {volume} {81}},\ \bibinfo {pages} {4816} (\bibinfo {year} {1998})},\ \Eprint
  {http://arxiv.org/abs/hep-ph/9806219} {arXiv:hep-ph/9806219 [hep-ph]}
  \BibitemShut {NoStop}%
\bibitem [{\citenamefont {Staig}\ and\ \citenamefont
  {Shuryak}(2011)}]{Staig:2010pn}%
  \BibitemOpen
  \bibfield  {author} {\bibinfo {author} {\bibfnamefont {P.}~\bibnamefont
  {Staig}}\ and\ \bibinfo {author} {\bibfnamefont {E.}~\bibnamefont
  {Shuryak}},\ }\href {\doibase 10.1103/PhysRevC.84.034908} {\bibfield
  {journal} {\bibinfo  {journal} {Phys. Rev. C}\ }\textbf {\bibinfo {volume}
  {84}},\ \bibinfo {pages} {034908} (\bibinfo {year} {2011})},\ \Eprint
  {http://arxiv.org/abs/1008.3139} {arXiv:1008.3139 [nucl-th]} \BibitemShut
  {NoStop}%
\bibitem [{\citenamefont {Lacey}\ \emph {et~al.}(2013)\citenamefont {Lacey},
  \citenamefont {Gu}, \citenamefont {Gong}, \citenamefont {Reynolds},
  \citenamefont {Ajitanand}, \citenamefont {Alexander}, \citenamefont {Mwai},\
  and\ \citenamefont {Taranenko}}]{Lacey:2013is}%
  \BibitemOpen
  \bibfield  {author} {\bibinfo {author} {\bibfnamefont {R.~A.}\ \bibnamefont
  {Lacey}}, \bibinfo {author} {\bibfnamefont {Y.}~\bibnamefont {Gu}}, \bibinfo
  {author} {\bibfnamefont {X.}~\bibnamefont {Gong}}, \bibinfo {author}
  {\bibfnamefont {D.}~\bibnamefont {Reynolds}}, \bibinfo {author}
  {\bibfnamefont {N.}~\bibnamefont {Ajitanand}}, \bibinfo {author}
  {\bibfnamefont {J.}~\bibnamefont {Alexander}}, \bibinfo {author}
  {\bibfnamefont {A.}~\bibnamefont {Mwai}}, \ and\ \bibinfo {author}
  {\bibfnamefont {A.}~\bibnamefont {Taranenko}},\ }\href@noop {} {\  (\bibinfo
  {year} {2013})},\ \Eprint {http://arxiv.org/abs/1301.0165} {arXiv:1301.0165
  [nucl-ex]} \BibitemShut {NoStop}%
\bibitem [{\citenamefont {Shuryak}\ and\ \citenamefont
  {Torres-Rincon}(2019)}]{Shuryak:2018lgd}%
  \BibitemOpen
  \bibfield  {author} {\bibinfo {author} {\bibfnamefont {E.}~\bibnamefont
  {Shuryak}}\ and\ \bibinfo {author} {\bibfnamefont {J.~M.}\ \bibnamefont
  {Torres-Rincon}},\ }\href {\doibase 10.1103/PhysRevC.100.024903} {\bibfield
  {journal} {\bibinfo  {journal} {Phys. Rev.}\ }\textbf {\bibinfo {volume}
  {C100}},\ \bibinfo {pages} {024903} (\bibinfo {year} {2019})},\ \Eprint
  {http://arxiv.org/abs/1805.04444} {arXiv:1805.04444 [hep-ph]} \BibitemShut
  {NoStop}%
\bibitem [{\citenamefont {Shuryak}\ and\ \citenamefont
  {Torres-Rincon}(2020{\natexlab{a}})}]{Shuryak:2019ikv}%
  \BibitemOpen
  \bibfield  {author} {\bibinfo {author} {\bibfnamefont {E.}~\bibnamefont
  {Shuryak}}\ and\ \bibinfo {author} {\bibfnamefont {J.~M.}\ \bibnamefont
  {Torres-Rincon}},\ }\href {\doibase 10.1103/PhysRevC.101.034914} {\bibfield
  {journal} {\bibinfo  {journal} {Phys. Rev.}\ }\textbf {\bibinfo {volume}
  {C101}},\ \bibinfo {pages} {034914} (\bibinfo {year} {2020}{\natexlab{a}})},\
  \Eprint {http://arxiv.org/abs/1910.08119} {arXiv:1910.08119 [nucl-th]}
  \BibitemShut {NoStop}%
\bibitem [{\citenamefont {Stephanov}(2011)}]{1104.1627}%
  \BibitemOpen
  \bibfield  {author} {\bibinfo {author} {\bibfnamefont {M.~A.}\ \bibnamefont
  {Stephanov}},\ }\href {\doibase 10.1103/PhysRevLett.107.052301} {\bibfield
  {journal} {\bibinfo  {journal} {Phys. Rev. Lett.}\ }\textbf {\bibinfo
  {volume} {107}},\ \bibinfo {pages} {052301} (\bibinfo {year} {2011})},\
  \Eprint {http://arxiv.org/abs/1104.1627} {arXiv:1104.1627 [hep-ph]}
  \BibitemShut {NoStop}%
\bibitem [{\citenamefont {DeMartini}\ and\ \citenamefont
  {Shuryak}(2020)}]{DeMartini:2020hka}%
  \BibitemOpen
  \bibfield  {author} {\bibinfo {author} {\bibfnamefont {D.}~\bibnamefont
  {DeMartini}}\ and\ \bibinfo {author} {\bibfnamefont {E.}~\bibnamefont
  {Shuryak}},\ }\href@noop {} {\  (\bibinfo {year} {2020})},\ \Eprint
  {http://arxiv.org/abs/2007.04863} {arXiv:2007.04863 [nucl-th]} \BibitemShut
  {NoStop}%
\bibitem [{\citenamefont {Shuryak}(2006)}]{Shuryak:2006eb}%
  \BibitemOpen
  \bibfield  {author} {\bibinfo {author} {\bibfnamefont {E.}~\bibnamefont
  {Shuryak}},\ }\bibfield  {booktitle} {\emph {\bibinfo {booktitle} {{Critical
  point and onset of deconfinement. Proceedings, 3rd Conference, CPOD2006,
  Florence, Itlay, July 3-6, 2006}}},\ }\href {\doibase 10.22323/1.029.0026}
  {\bibfield  {journal} {\bibinfo  {journal} {PoS}\ }\textbf {\bibinfo {volume}
  {CPOD2006}},\ \bibinfo {pages} {026} (\bibinfo {year} {2006})},\ \Eprint
  {http://arxiv.org/abs/nucl-th/0609011} {arXiv:nucl-th/0609011 [nucl-th]}
  \BibitemShut {NoStop}%
\bibitem [{\citenamefont {Berges}\ \emph {et~al.}(2002)\citenamefont {Berges},
  \citenamefont {Tetradis},\ and\ \citenamefont {Wetterich}}]{Berges:2000ew}%
  \BibitemOpen
  \bibfield  {author} {\bibinfo {author} {\bibfnamefont {J.}~\bibnamefont
  {Berges}}, \bibinfo {author} {\bibfnamefont {N.}~\bibnamefont {Tetradis}}, \
  and\ \bibinfo {author} {\bibfnamefont {C.}~\bibnamefont {Wetterich}},\ }\href
  {\doibase 10.1016/S0370-1573(01)00098-9} {\bibfield  {journal} {\bibinfo
  {journal} {Phys. Rept.}\ }\textbf {\bibinfo {volume} {363}},\ \bibinfo
  {pages} {223} (\bibinfo {year} {2002})},\ \Eprint
  {http://arxiv.org/abs/hep-ph/0005122} {arXiv:hep-ph/0005122} \BibitemShut
  {NoStop}%
\bibitem [{\citenamefont {Dupuis}\ \emph {et~al.}(2020)\citenamefont {Dupuis},
  \citenamefont {Canet}, \citenamefont {Eichhorn}, \citenamefont {Metzner},
  \citenamefont {Pawlowski}, \citenamefont {Tissier},\ and\ \citenamefont
  {Wschebor}}]{Dupuis:2020fhh}%
  \BibitemOpen
  \bibfield  {author} {\bibinfo {author} {\bibfnamefont {N.}~\bibnamefont
  {Dupuis}}, \bibinfo {author} {\bibfnamefont {L.}~\bibnamefont {Canet}},
  \bibinfo {author} {\bibfnamefont {A.}~\bibnamefont {Eichhorn}}, \bibinfo
  {author} {\bibfnamefont {W.}~\bibnamefont {Metzner}}, \bibinfo {author}
  {\bibfnamefont {J.}~\bibnamefont {Pawlowski}}, \bibinfo {author}
  {\bibfnamefont {M.}~\bibnamefont {Tissier}}, \ and\ \bibinfo {author}
  {\bibfnamefont {N.}~\bibnamefont {Wschebor}},\ }\href@noop {} {\  (\bibinfo
  {year} {2020})},\ \Eprint {http://arxiv.org/abs/2006.04853} {arXiv:2006.04853
  [cond-mat.stat-mech]} \BibitemShut {NoStop}%
\bibitem [{\citenamefont {Tsypin}(1994)}]{hep-lat/9401034}%
  \BibitemOpen
  \bibfield  {author} {\bibinfo {author} {\bibfnamefont {M.~M.}\ \bibnamefont
  {Tsypin}},\ }\href@noop {} {\  (\bibinfo {year} {1994})},\ \Eprint
  {http://arxiv.org/abs/hep-lat/9401034} {arXiv:hep-lat/9401034 [hep-lat]}
  \BibitemShut {NoStop}%
\bibitem [{\citenamefont {Nonaka}\ and\ \citenamefont
  {Asakawa}(2005)}]{nucl-th/0410078}%
  \BibitemOpen
  \bibfield  {author} {\bibinfo {author} {\bibfnamefont {C.}~\bibnamefont
  {Nonaka}}\ and\ \bibinfo {author} {\bibfnamefont {M.}~\bibnamefont
  {Asakawa}},\ }\href {\doibase 10.1103/PhysRevC.71.044904} {\bibfield
  {journal} {\bibinfo  {journal} {Phys. Rev.}\ }\textbf {\bibinfo {volume}
  {C71}},\ \bibinfo {pages} {044904} (\bibinfo {year} {2005})},\ \Eprint
  {http://arxiv.org/abs/nucl-th/0410078} {arXiv:nucl-th/0410078 [nucl-th]}
  \BibitemShut {NoStop}%
\bibitem [{\citenamefont {Adamczewski-Musch}\ \emph {et~al.}(2020)\citenamefont
  {Adamczewski-Musch} \emph {et~al.}}]{Adamczewski-Musch:2020slf}%
  \BibitemOpen
  \bibfield  {author} {\bibinfo {author} {\bibfnamefont {J.}~\bibnamefont
  {Adamczewski-Musch}} \emph {et~al.} (\bibinfo {collaboration} {HADES}),\
  }\href {\doibase 10.1103/PhysRevC.102.024914} {\bibfield  {journal} {\bibinfo
   {journal} {Phys. Rev. C}\ }\textbf {\bibinfo {volume} {102}},\ \bibinfo
  {pages} {024914} (\bibinfo {year} {2020})},\ \Eprint
  {http://arxiv.org/abs/2002.08701} {arXiv:2002.08701 [nucl-ex]} \BibitemShut
  {NoStop}%
\bibitem [{\citenamefont {Adam}\ \emph {et~al.}(2020)\citenamefont {Adam} \emph
  {et~al.}}]{Adam:2020unf}%
  \BibitemOpen
  \bibfield  {author} {\bibinfo {author} {\bibfnamefont {J.}~\bibnamefont
  {Adam}} \emph {et~al.} (\bibinfo {collaboration} {STAR}),\ }\href@noop {} {\
  (\bibinfo {year} {2020})},\ \Eprint {http://arxiv.org/abs/2001.02852}
  {arXiv:2001.02852 [nucl-ex]} \BibitemShut {NoStop}%
\bibitem [{\citenamefont {Shuryak}\ and\ \citenamefont
  {Torres-Rincon}(2020{\natexlab{b}})}]{Shuryak:2020yrs}%
  \BibitemOpen
  \bibfield  {author} {\bibinfo {author} {\bibfnamefont {E.}~\bibnamefont
  {Shuryak}}\ and\ \bibinfo {author} {\bibfnamefont {J.~M.}\ \bibnamefont
  {Torres-Rincon}},\ }\href@noop {} {\  (\bibinfo {year}
  {2020}{\natexlab{b}})},\ \Eprint {http://arxiv.org/abs/2005.14216}
  {arXiv:2005.14216 [nucl-th]} \BibitemShut {NoStop}%
\bibitem [{\citenamefont {Zhao}\ \emph {et~al.}(2020)\citenamefont {Zhao},
  \citenamefont {Shen}, \citenamefont {Ko}, \citenamefont {Liu},\ and\
  \citenamefont {Song}}]{Zhao:2020irc}%
  \BibitemOpen
  \bibfield  {author} {\bibinfo {author} {\bibfnamefont {W.}~\bibnamefont
  {Zhao}}, \bibinfo {author} {\bibfnamefont {C.}~\bibnamefont {Shen}}, \bibinfo
  {author} {\bibfnamefont {C.~M.}\ \bibnamefont {Ko}}, \bibinfo {author}
  {\bibfnamefont {Q.}~\bibnamefont {Liu}}, \ and\ \bibinfo {author}
  {\bibfnamefont {H.}~\bibnamefont {Song}},\ }\href {\doibase
  10.1103/PhysRevC.102.044912} {\bibfield  {journal} {\bibinfo  {journal}
  {Phys. Rev. C}\ }\textbf {\bibinfo {volume} {102}},\ \bibinfo {pages}
  {044912} (\bibinfo {year} {2020})},\ \Eprint
  {http://arxiv.org/abs/2009.06959} {arXiv:2009.06959 [nucl-th]} \BibitemShut
  {NoStop}%
\bibitem [{\citenamefont {E.~Brezin}\ and\ \citenamefont
  {Zinn-Justin}(1976)}]{Zinn-Justin}%
  \BibitemOpen
  \bibfield  {author} {\bibinfo {author} {\bibfnamefont {J.~C. L.~G.}\
  \bibnamefont {E.~Brezin}}\ and\ \bibinfo {author} {\bibfnamefont
  {J.}~\bibnamefont {Zinn-Justin}},\ }in\ \href@noop {} {\emph {\bibinfo
  {booktitle} {Phase Transition and Critical Phenomena}}},\ Vol.~\bibinfo
  {volume} {6},\ \bibinfo {editor} {edited by\ \bibinfo {editor} {\bibfnamefont
  {C.}~\bibnamefont {Domb}}\ and\ \bibinfo {editor} {\bibfnamefont {M.~S.}\
  \bibnamefont {Green}}}\ (\bibinfo  {publisher} {Academic Press},\ \bibinfo
  {address} {New York},\ \bibinfo {year} {1976})\BibitemShut {NoStop}%
\end{thebibliography}%

\end{document}